\documentclass{jfm}

\usepackage{natbib}
\usepackage{graphicx}

\ifCUPmtlplainloaded \else
  \checkfont{eurm10}
  \iffontfound
    \IfFileExists{upmath.sty}
      {\typeout{^^JFound AMS Euler Roman fonts on the system,
                   using the 'upmath' package.^^J}%
       \usepackage{upmath}}
      {\typeout{^^JFound AMS Euler Roman fonts on the system, but you
                   dont seem to have the}%
       \typeout{'upmath' package installed. JFM.cls can take advantage
                 of these fonts,^^Jif you use 'upmath' package.^^J}%
      }
  \else
  \fi
\fi

\ifCUPmtlplainloaded \else
  \checkfont{msam10}
  \iffontfound
    \IfFileExists{amssymb.sty}
      {\typeout{^^JFound AMS Symbol fonts on the system, using the
                'amssymb' package.^^J}%
       \usepackage{amssymb}%
       \let\le=\leqslant  \let\leq=\leqslant
       \let\ge=\geqslant  
      }{}
  \fi
\fi

\ifCUPmtlplainloaded \else
  \IfFileExists{amsbsy.sty}
    {\typeout{^^JFound the 'amsbsy' package on the system, using it.^^J}%
     \usepackage{amsbsy}}
    {\providecommand\boldsymbol[1]{\mbox{\boldmath $##1$}}}
\fi

\newsavebox{\astrutbox}
\sbox{\astrutbox}{\rule[-5pt]{0pt}{20pt}}

\def\bs{\boldsymbol}
\def\gsim{\;\rlap{\lower 2.5pt
\hbox{$\sim$}}\raise 1.5pt\hbox{$>$}\;}
\def\lsim{\;\rlap{\lower 2.5pt
\hbox{$\sim$}}\raise 1.5pt\hbox{$<$}\;}
\newlength{\myfigwidth}
\title[Particle Relative Velocity]{Relative Velocity of Inertial Particles in Turbulent Flows}
\author[L. Pan \& P. Padoan]{L\ls I\ls U\ls B\ls I\ls N\ns P\ls A\ls N$^{1,2}$\thanks{Email address for correspondence: liubin.pan@asu.edu}\ns
\and P\ls A\ls O\ls L\ls O\ns  P\ls A\ls D\ls O\ls A\ls N$^{1,3}$}

\affiliation{$^1$Department of Physics, University of California, San Diego, 
CASS/UCSD 0424, 9500 Gilman Drive, La Jolla, CA 92093-0424 \\
$^2$School of Earth and Space Exploration, Arizona State University, 
P.O. Box 871404, Tempe, AZ 85287-1404\\
$^3$ICREA \& ICC, University of Barcelona, Marti i Franqu\`{e}s 1, E-08028 
Barcelona, Spain}

\pubyear{??}
\volume{??}
\pagerange{??}
\date{?? and in revised form ??}
\begin{document}
\maketitle

\begin{abstract}
We present a model for the relative velocity of inertial
particles in turbulent flows that provides new physical 
insight into this problem. Our general formulation shows 
that the relative velocity has contributions from two 
terms, referred to as the {\it generalized acceleration} 
and {\it generalized shear} terms, because they reduce 
to the well known acceleration and shear terms in the 
Saffman-Turner limit. 
The generalized shear term represents 
particles' memory of the flow velocity 
difference along their trajectories and 
depends on the inertial particle pair dispersion 
backward in time. The importance of this backward 
dispersion in determining the particle relative velocity 
is emphasized. 
We find that our model with a two-phase separation 
behavior, an early ballistic  phase and a later tracer-like 
phase, as found by recent simulations for the forward (in time) 
dispersion of inertial particle pairs, gives good fits 
to the measured relative speeds from simulations 
at low Reynolds numbers. In the monodisperse case 
with identical particles, the generalized acceleration term 
vanishes and the relative velocity is determined by the 
generalized shear term. At large Reynolds numbers,  
our model gives a $St^{1/2}$ dependence of the 
relative velocity on the Stokes number $St$ in 
the inertial range for both the ballistic behavior 
and the Richardson separation law. 
This leads to the same inertial-range scaling 
for the two-phase separation that well fits the simulation 
results.  Our calculations for the bidisperse case show that, 
with the friction timescale of one particle fixed, the relative 
speed as a function of the other particle's friction time 
has a dip when the two timescales are similar. 
This indicates that similar-size particles tend to 
have stronger velocity correlation than different ones. 
We find that the primary contribution at the dip, i.e., 
for similar particles, is from the generalized shear term, 
while the generalized acceleration term is dominant for particles of very 
different sizes. Future numerical studies are motivated to 
check the accuracy of the assumptions made in our model 
and to investigate the backward-in-time dispersion of inertial 
particle pairs in turbulent flows.   
\end{abstract}


\section{Introduction}
The dynamics of inertial particles suspended in turbulent flows is of both theoretical 
and practical importance. Its applications range from industrial processes, e.g., 
turbulent spray combustion, aerosols and raindrop formation in terrestrial clouds, 
and dust grain dynamics in astrophysical environments such as interstellar media, 
protoplanetary disks, and planetary atmospheres. Particle collisions in a turbulent 
flow are of particular interest because they determine the growth of the particles 
by coagulation. The estimate of the collision rate requires the understanding of 
two interesting phenomena regarding inertial particles in turbulence, the preferential 
concentration and the turbulence-induced relative velocity. The latter is the focus 
of the present work.   

Our primary motivations for this study is its application to theoretical 
modeling of planetesimal formation in protoplanetory disks.  
Planetesimals are objects of kilometer size that can further 
grow into fully-fledged planets.  The formation of these objects is 
believed to start from the growth of dust grains of micrometer 
size by collisional coagulation \cite[e.g.,][]{wei80}.   
Particles involved in this process have an extensive size range, 
from micrometer to kilometer,  which corresponds to a 
range of friction timescale that covers all the scales 
(from the dissipation range, the inertial range to the 
outer scales) in the turbulence in protoplanetary disks.   
Therefore, a detailed understanding of particle collision 
velocities for a whole range of particle sizes induced by 
turbulence motions is crucial to investigating particle growth in 
these disks.  Dust grain collisions in protoplanetary disks 
do not always lead to coagulation. The grains become less 
sticky as the size increases. If the relative speed is large, 
the collision between two large particles may cause them 
to fragment or to simply bounce off each other \cite[e.g.,] []{blu08}. 
To judge the outcome of collisions between large particles thus
requires an accurate prediction for the collision speed. 

Besides turbulence, there are other effects that can 
induce relative velocities between particles. For example,  
gravity gives rise to differential settling for particles of different 
sizes, which can have important contribution to the relative 
speed between these particles. In astrophysical environments, 
radiation pressure and the coupling with magnetic fields 
(through electric charge on the grain surface) can also 
play an important role. In the present study, we will neglect 
these effects, and concentrate on the turbulence-induced relative velocity. 
We aim at a physical understanding of relative velocities 
induced by turbulent motions alone, which is clearly a crucial step 
toward an accurate model for the particle collisions  
in the presence of both turbulence and the other complexities. 
The model presented here could be extended to take the 
other effects into account.  Although our work is motivated 
by the problem of dust grain collisions in astrophysical 
environments, it has applications in other contexts such 
as droplet formation in cloud physics \cite[e.g.,] []{saf56}.   


The relative velocity of two nearby particles induced by turbulent motions 
has been extensively studied. \cite{saf56} considered particles with 
small inertia in the limit that the friction time, $\tau_p$, of 
both particles is much smaller than the Kolmogorov timescale, 
$\tau_{\eta}$. 
This limit is also expressed as $St \ll 1$, where the Stokes 
number, $St$, is defined as the ratio 
of $\tau_p$ to $\tau_{\eta}$. In this limit, the particle velocity at a given point 
can be approximately obtained 
from the (1st-order) Taylor expansion of the particle momentum equation 
(eq. (2.1) below). For two particles with a separation, $r$, much smaller than the 
Kolmogorov scale, $\eta$, \cite{saf56} derived a result for the 
average radial relative velocity, $\langle|w_r|\rangle$. In the absence of 
gravity, there are two terms that contribute to the relative speed, a shear 
term and an acceleration term \cite[see][]{aya08}, 
\begin{equation}
\langle|w_r|\rangle = \sqrt{\frac{2}{\pi}} \left (\frac{1}{15} \frac {\bar{\epsilon}}{\nu} r^2 + 
(\tau_{p2}-\tau_{p1} )^2 a^2 \right)^{1/2}       
\end{equation} 
where $\bar{\epsilon}$ and $\nu$ are, respectively, the average dissipation rate and the 
kinematic viscosity of the turbulent flow, $\tau_{p1}$ and $\tau_{p2}$ are the 
friction timescales of the two particles, and $a$ is the rms acceleration, i.e., 
$a^2 =\left \langle \left(\frac{Du}{Dt}\right)^2 \right\rangle$.   
The factor $\sqrt{2/\pi}$ is from the conversion of the radial relative velocity 
variance, $\langle w_r^2 \rangle$, to $\langle|w_r|\rangle$ assuming a 
Gaussian distribution for $w_r$. Note that the shear term is from the 
longitudinal structure function, $S_{ll}(r)$,  of the flow, which is given 
by $\frac{1}{15}\frac{\bar{\epsilon}}{\nu} r^2 $ for $ r \lsim \eta$.  
In the Saffman-Turner (S-T) limit, the particle velocity follows the local flow 
velocity very closely, thus the velocity of the two particles at a 
distance $r<\eta$ is highly correlated because of the strong flow 
velocity correlation across a small distance.          

The opposite limit is that of heavy particles with friction timescales much 
larger than the Lagrangian correlation timescale, $T_L$. In this limit, 
the velocities of two particles coming together are essentially uncorrelated. 
This is because particles with $\tau_p \gg T_L$ have long-time memory,  
and their current velocities have substantial contributions from the flow 
velocities on their trajectories in the past. These flow velocities at early times 
are likely to be uncorrelated because the particles were far away from each other.  
The relative velocity in this limit is thus determined by the sum of the velocity 
variances of the two particles, 
\begin{equation}
\langle|w_r|\rangle = \sqrt{\frac{2}{\pi}} \left [ \left(v'^{(1)}\right)^2 + \left (v'^{(2)} \right)^2 \right ]^{1/2}
\end{equation} 
where $v'^{(1)}$ and $v'^{(2)}$ denote the particle rms velocities.  
In the limit $\tau_p \gg T_L$, they are given by \cite[e.g.,][]{abr75},   
\begin{equation}
\left(v'^{(1)}\right)^2  \simeq u'^2 \frac{T_L}{\tau_{p1}} {\rm ;} \hspace{1cm} \left(v'^{(2)}\right)^2 \simeq u'^2 \frac{T_L}{\tau_{p2}} 
\end{equation}  
where $u'$ is the rms of the flow velocity fluctuations.  In the derivation of 
eq. (1.3), the temporal correlation of the flow velocity on a particle's trajectory 
is approximated by the Lagrangian correlation for tracer particles. 
Discussions on the validity of this assumption will be given in \S 2.2.

The problem of the relative velocity of inertial particles in these two extreme 
limits is physically clear, and the results given above are easy to understand 
and are expected to be generally robust. On the other hand, for particles 
with intermediate inertia, $\tau_{\eta} \lsim \tau_p \lsim T_L$, the problem is more 
complicated and is less well understood. The velocities of two 
nearby particles with intermediate $\tau_p$ are partially 
correlated and the degree of correlation, intermediate
between the two limits, is not easy to evaluate. We will 
point out that a very important factor in determining this 
correlation is the distance between the trajectories of the 
two particles as a function of time before they come close to each other. 
To our knowledge, this point has not been clearly recognized 
or explicitly emphasized in the literature. We will show how 
the separation of two nearby inertial particles backward in 
time affects the relative velocity between particles with 
intermediate inertia, $\tau_{\eta} \lsim \tau_p \lsim T_L$. 

A successful theory for particles of all sizes needs to correctly predict and 
explain the behavior of the relative velocity between particles with intermediate 
friction time, as well as recover the results in the two extreme limits. 
An example of particular theoretical interest is that of identical particles, 
referred to as the monodisperse case. In the S-T limit ($\tau_p \ll \tau_\eta$), 
the acceleration term in eq. (1.1) vanishes for the monodisperse case and 
the relative velocity does not depend on the friction time. It is constant at a given 
distance, $r$, and increases linearly with $r$. In the opposite limit with 
$\tau_p \gg T_L$, the relative velocity decreases with the friction time as $\tau_p^{-1/2}$ or 
$St^{-1/2}$, according to eqs. (1.2) and (1.3). The question of how $\langle |w_r| \rangle$ 
scales with $\tau_p$ for $\tau_p$ in a range corresponding to the inertial range 
of the turbulent flow, and how it connects with the two extreme limits has not 
been systematically studied or fully understood. This is one of the primary 
goals of the present paper. We find that the dispersion of particles backward in time is 
crucial to answer this question.  

The existing models have very different predictions for the relative speed in the 
inertial range for the monodisperse case \cite[][]{vol80, wil83, yuu84, kru97, zai03, zai06, aya08}. 
A detailed discussion of the qualitative differences between these models and 
their problems will be given in \S 4.  An important reason for the problems in most 
of the previous models is that they did not clearly recognize or carefully consider 
the effect of the particle pair separation backward in time (except for the 
differential model by \cite{zai03, zai06} to be discussed below, which we think has the 
particle backward separation implicitly included). The role of this backward 
separation will be revealed along the formulation of our model.

    
In the previous studies that cover a whole range of Stokes numbers, the 
differential model by Zaichik and collaborators \cite[][]{zai03a, zai03, zai06} is 
perhaps the most complete one, as it examines the effect of preferential 
clustering and the relative speed simultaneously. We will refer to this model 
as Zaichik {\it et al}.'s 
model or the model by Zaichik {\it et al}. Assuming Gaussian statistics for the 
flow velocity, the model first sets up an equation for the joint probability distribution 
function (pdf) of the particle separation and the relative velocity.  
Deriving the first 3 moment equations of the pdf equation and closing these moment 
equations by a quasi-normal approximation, Zaichik {\it et al}. were able to obtain a 
set of partial differential equations for the particle density correlation function 
(usually referred to as the radial distribution function) 
and the particle velocity structure functions. The solution of 
the differential equations reproduces the two extreme limits discussed above 
and predicts that the relative velocity of identical particles with intermediate inertia 
is proportional to $\tau_p^{1/2}$ or, equivalently, $St^{1/2}$. The validity 
of this prediction remains to be confirmed by high-resolution simulations. 
Despite the elegant mathematical formulation and good agreement with results of 
direct numerical simulations of turbulence with low Reynolds numbers, the model 
lacks physical transparency in its approximations, especially the quasi-normal assumption.
One of the goals of our model is to elucidate the physics behind the 
inertial range scaling of the relative velocity.


\cite{fal02} found that an effect, named the sling effect, 
has a significant contribution to the relative speed 
\cite[see also][]{wil05, wil06, fal07}.  The physical picture 
of the effect is that,  at regions with large negative velocity gradients, 
faster moving particles can catch up the slower ones from behind \cite[][]{fal02},  leading to 
trajectory crossing of the particles (\cite{bec05}, also see Fig. 1 in 
\cite{fal07} for an illustration). This results in a larger relative speed 
than the S-T prediction for small particles. \cite{fal07} showed that 
the effect starts to be important  for $St \gsim 0.2$ and gives a relative 
speed several time larger than eq (1.1) for $St$ between $0.2$ and $1$.  
We will point out a common element shared by the sling effect and 
our model: the contribution to the relative speed from the particles' 
memory of the flow velocity difference in the past.  
       
The paper is organized as follows. We present the formulation of our model in \S 2 
(a general formulation in \S2.1, and basic assumptions and approximations in 
\S2.2 and \S 2.3). The results for identical particles (monodisperse) and different 
particles (bidisperse) are given in \S3.1 and \S 3.2, respectively.  In \S4, we discuss 
previous models and compare them with our model. Conclusions are given in \S 5.  
  
\section{The Model}

\subsection{The General Formulation}
The velocity, ${\bs v}(t)$, of a particle with friction time, $\tau_p$, 
can be obtained by integrating the momentum equation, 
\begin{equation}
\frac {d {\bs v} } {dt} = \frac { {\bs u} \left( {\bs X} (t), t \right) - {\bs v}} {\tau_p}            
\end{equation}
where $\bs u ({\bs x}, t)$ denotes the flow velocity field and ${\bs X} (t)$ is the 
position of the particle at time $t$. Clearly, ${\bs u} \left( {\bs X} (t), t \right)$ is the 
flow velocity at the positions of the particle along its trajectory (we will refer to it
also as the flow velocity ``seen" by the particle). 
The particle trajectory is given by,  
\begin{equation}
{\bs X} (t) = {\bs X}_0 + \int_{t_0}^{t} {\bs v} (t') dt'  
\end{equation} 
where ${\bs X}_0$ is the particle position at a given time $t_0$.

Equation (2.1) has a formal solution, 
\begin{equation}
{\bs v}(t) = {\bs v}_0 \exp \left(- \frac{t - t_0}{\tau_p} \right) + \frac {1} {\tau_p }
\int_{t_0}^t  {\bs u} \left({\bs X} (\tau), \tau \right) \exp \left(- \frac{t-\tau}{\tau_p}\right) 
d\tau           
\end{equation}
where $\bs {v}_0$ is the particle velocity at $t_0$. 

We are interested in deriving the relative velocity between two particles at a 
distance ${\bs r}$ at a given time $t$. We label the two particles by superscripts 
``(1)" and ``(2)". For example, their velocities  at $t$ are denoted as 
${\bs v}^{(1)}(t)$ and ${\bs v}^{(2)}(t)$, respectively.  
To evaluate the average relative speed, we will calculate the velocity 
structure tensor, $S_{{\rm p}ij}$, of the two particles, 
\begin{equation}
S_{{\rm p}ij} ({\bs r}, t) =\left \langle \left(v^{(1)}_i -v^{(2)}_i\right) \left(v^{(1)}_j -v^{(2)}_j\right) \right \rangle           
\end{equation}  
where $\langle \cdot \cdot \cdot \rangle$ denotes the ensemble average. 
The particle velocities can be solved by integrating equation (2.3) and the trajectories 
of the two particles are subject to a constraint,    
\begin{equation}
{\bs X}^{(1)}(t) - {\bs X}^{(2)}(t) = {\bs r}.   
\end{equation}    
which means that two particles happen to be separated by ${\bs r}$ at $t$.  
We will particularly consider small values of $r$ (below $\eta$) because we are 
interested in the collision speed, which is essentially the relative velocity of the 
two particles over a distance of the particle size. From the structure tensor $S_{{\rm p}ij}$, 
we will obtain the longitudinal structure function $S_{{\rm p}ll}$, which, by definition, 
is the radial relative velocity variance, $\langle w_r^2 \rangle$. 
Although only small $r$ will be considered in the paper, our model 
can predict the structure function at all separations.  By a comparison with their results, 
our model may provide an explanation for the inertial particle structure functions     
found in \cite{bec09b}.  
 

The particle structure tensor can be written as,  
\begin{equation}  
S_{{\rm p} ij} = \left\langle v^{(1)}_i v^{(1)}_j \right\rangle - \left\langle v^{(1)}_i v^{(2)}_j \right\rangle -  
\left\langle v^{(2)}_i v^{(1)}_j \right\rangle + \left\langle v^{(2)}_i v^{(2)}_j \right\rangle. 
\end{equation}  
Note that the cross terms correspond to the particle velocity correlations discussed 
in \S 1, where it was argued that a careful treatment of such correlations is essential 
for modeling the relative velocity of particles with intermediate inertia.   
 
To calculate the structure tensor, we insert eq. (2.3) for the particle velocities into eq. (2.6). 
For simplicity in notations, we will set the time when the particle relative speed is measured 
(i.e., $t$ in eq. (2.4)) to be zero, 
and assume it is far from the initial time (since we are interested in the relative velocity 
for a statistically stationary state). This allows us to set $t_0$ in eq. (2.3) to $-\infty$.  

We analyze the four terms in eq. (2.6) one by one. The first term on the r.h.s. corresponds to 
the velocity variance of particle (1).  For this term, only the velocity of particle (1) is involved and we have,   
\begin{equation}
\left \langle v^{(1)}_i v^{(1)}_j \right \rangle  =  \int_{-\infty}^0  \frac {d\tau}{\tau_{p1}} \int_{-\infty}^0 \frac {d\tau'}{\tau_{p1}}
 \left \langle u_i^{(1)} (\tau) u_j^{(1)}(\tau') \right \rangle \exp \left( \frac {\tau}{\tau_{p1}} \right) \exp \left (\frac{\tau'}{\tau_{p1}}\right) 
\end{equation}
where $u_i^{(1)} (t)=u_i \left({\bs X}^{(1)} (t), t\right)$ denotes the flow velocity on the trajectory of 
particle (1). The exponential factors here represent the memory loss of the particles.  
The integral limits in eq. (2.7) (see also eq. (2.9)) suggest that it is the flow velocity 
the particles saw in the past that is relevant in determining the particles' velocities 
at the current time. The relative position of the two particles back in time will 
be shown to play an important role in the prediction of their relative velocity.   
We will call $\left \langle u_i^{(1)} (\tau) u_j^{(1)}(\tau') \right \rangle$ in the integral the trajectory correlation 
tensor and denote it as $B^{(1)}_{{\rm T}ij}$, i.e., 
\begin{equation}
B^{(1)}_{{\rm T}ij}(\tau,\tau') =\left \langle u_i^{(1)} (\tau) u_j^{(1)}(\tau') \right \rangle
\end{equation} 
where the subscript ``T" stands for ``trajectory". 

The result for the 4th term on the rhs of eq. (2.6) is similar. One only needs to replace $\tau_{p1}$ in eq. (2.7) by $\tau_{p2}$,  
and $B^{(1)}_{{\rm T}ij}$ by $B^{(2)}_{{\rm T}ij} \equiv \left \langle u_i^{(2)} (\tau) u_j^{(2)}(\tau') \right \rangle$. 
If the two particles are identical and have the same friction time, $\left \langle v^{(1)}_i v^{(1)}_j \right\rangle$  
is 
equal to $\left\langle v^{(2)}_i v^{(2)}_j \right\rangle$. 
These two terms correspond to the velocity variance of each particle and will be called the 
velocity variance terms.   

The exact form of $B_{{\rm T}ij}$ as a function of the friction time is not available. 
In the limit of vanishing $\tau_p$ (i.e., passive tracers),
this correlation tensor would approach the Lagrangian correlation tensor, $B_{{\rm L}ij}$, 
of the flow, which has been extensively studied \cite[e.g.,][]{yeu89}.
A common approximation is to set $B_{{\rm T}ij}$ equal to $B_{{\rm L}ij}$ for particles with any 
$\tau_p$ \cite[e.g.,][]{zai03a, zai03, zai06, aya08}. Physically, it corresponds to the assumption 
that the trajectory of any inertial particle is not far away from that of a tracer particle starting from 
the same initial condition. We will adopt this assumption in our calculations and its validity will 
be discussed in \S 2.2. 

The cross correlation terms in eq. (2.6) can be evaluated with the same approach. The second term on the r.h.s. is given by,  
\begin{equation}
\left \langle v^{(1)}_i v^{(2)}_j \right \rangle  =  \int_{-\infty}^0  \frac {d\tau}{\tau_{p1}}\int_{-\infty}^0 \frac{d\tau'}{\tau_{p2}}
\left \langle u_i^{(1)} (\tau) u_j^{(2)}(\tau') \right \rangle \exp \left(\frac{\tau}{\tau_{p1}} 
\right) \exp \left(\frac{\tau'}{\tau_{p2}}\right).   
\label{eq1}
\end{equation}
The result for the term $\left \langle v^{(2)}_i v^{(1)}_j \right\rangle$ in eq. (2.6) is similar to 
eq. (2.9). 
The sum of these two 
tensors can be written as,  
\begin{equation}
\left \langle u_i^{(1)} (\tau) u_j^{(2)}(\tau') \right \rangle + \left \langle u_i^{(2)} (\tau) u_j^{(1)}(\tau') \right \rangle 
= B^{(1)}_{{\rm T}ij}(\tau, \tau') + B^{(2)}_{{\rm T}ij}(\tau, \tau') - S_{{\rm T}ij}({\bs r}; \tau, \tau') 
\end{equation} 
where 
the tensor $S_{{\rm T}ij}$ is defined as 
\begin{equation}
S_{{\rm T}ij} ({\bs r}; \tau, \tau') = \left \langle \left[u_i^{(1)} (\tau) - u_i^{(2)} (\tau) \right] \left[ u_j^{(1)} (\tau') - u_j^{(2)} (\tau') \right]\right \rangle. 
\end{equation} 
Clearly, $S_{{\rm T}ij}$ is the correlation of the flow velocity difference at the positions of the
particles on their trajectories at times $\tau$ and $\tau'$. We have explicitly indicated 
the dependence of the tensor on the particle separation at time zero. The ensemble average 
on the r.h.s. includes an average over the probability distribution of the flow velocity difference 
at time zero. By analogy with $B_{{\rm T}ij} $, we will call $S_{{\rm T}ij}$ the trajectory 
structure tensor. 

Since  $S_{{\rm T}ij}$ has not been directly studied, we will give an approximate estimate 
for  it  in \S2.3.  
For example, we will relate the flow velocity difference, ${\bs u}^{(1)} (t) - {\bs u}^{(2)}(t)$, 
along the trajectories by the two particles, to the separation of the two particles at $t$, 
assuming the velocity difference scaling in the Eulerian frame applies to 
the velocity difference on the particles' trajectories. The uncertainty in this assumption will be discussed in \S2.3.  We will denote the 
particle separation at a given time $t$ as ${\bs \rho}(t)$, which, in our notation,
is given by ${\bs X}^{(1)}(t) - {\bs X}^{(2)} (t)$.  Note that ${\bs \rho}$ is a stochastic
vector because of the particle dispersion by turbulent motions. 





Combining eqs. (2.6), (2.7), (2.8), (2.9) and (2.10), we finally arrive at the formula for the 
velocity structure tensor of two particles separated by ${\bs r}$, 
\begin{equation}
S_{{\rm p}ij} ({\bs r}) = \mathcal{A}_{ij} +\mathcal{D}_{ij} {\rm ,} 
\end{equation}
where
\begin{equation}
\begin{array}{lll}
\mathcal{A}_{ij} = \hspace{2mm}{\displaystyle \int_{-\infty}^0  \frac {d\tau}{\tau_{p1}} \int_{-\infty}^0 \frac {d\tau'}{\tau_{p1}} B^{(1)}_{{\rm T}ij}(\tau, \tau') \exp \left (\frac {\tau}{\tau_{p1}} \right) \exp \left(\frac{\tau'}{\tau_{p1}}\right)} \\ 
\hspace{0.9cm} {\displaystyle -\int_{-\infty}^0  \frac {d\tau}{\tau_{p1}} \int_{-\infty}^0 \frac {d\tau'}{\tau_{p2}} \left( B^{(1)}_{{\rm T}ij}(\tau, \tau') + B^{(2)}_{{\rm T}ij}(\tau, \tau') \right) \exp \left (\frac {\tau}{\tau_{p1}} \right) \exp \left(\frac{\tau'}{\tau_{p2}}\right) }\\ 
\hspace{0.9cm}  {\displaystyle  + \int_{-\infty}^0  \frac {d\tau}{\tau_{p2}}\int_{-\infty}^0 \frac{d\tau'}{\tau_{p2}}  B^{(2)}_{{\rm T}ij}(\tau, \tau')  
\exp \left(\frac{\tau}{\tau_{p2}}\right) \exp \left(\frac{\tau'}{\tau_{p2}}\right) }
\end{array}
\end{equation}
and
\begin{equation}            
\mathcal{D}_{ij} = \int_{-\infty}^0  \frac {d\tau}{\tau_{p1}}\int_{-\infty}^0 \frac{d\tau'}{\tau_{p2}}
S_{{\rm T}ij} ({\bs r}; \tau, \tau') \exp \left(\frac{\tau}{\tau_{p1}} \right) \exp \left(\frac{\tau'}{\tau_{p2}}\right).
\end{equation} 
In the trajectory structure tensor in $\mathcal{D}_{ij}$,  the dependence on $\bs{r}$ is 
from the requirement  that particle separation $\bs{\rho}$  is equal to ${\bs r}$ at time zero, i.e.,   
\begin{equation}
{\bs \rho}(0) = {\bs r}.   
\end{equation} 


     
The result, eq. (2.12), is written in such a way that the first term $\mathcal{A}_{ij}$ only 
depends on the 1-particle trajectory correlation tensor and the second term 
$\mathcal{D}_{ij}$ only on the 2-particle trajectory structure tensor. There are 
also physical reasons to split $S_{{\rm p}ij}$ into these two terms. First, the $\mathcal{A}_{ij}$ 
term vanishes for identical particles with $\tau_{p1}=\tau_{p2}$, and only $\mathcal{D}_{ij}$ 
contributes to the relative speed in the monodisperse case. On the other hand, for particles of 
very different sizes, $\mathcal{A}_{ij}$ dominates the contribution to the relative speed (see \S 3.2). 
Second, in the S-T limit, $\mathcal{A}_{ij}$ and $\mathcal{D}_{ij}$ reduce to the acceleration
term and the shear term in eq. (1.1), respectively.  Therefore, our formulation can 
be regarded as one that extends eq. (1.1) from the low-inertia limit to the whole 
range of particle sizes. We will refer to $\mathcal{A}_{ij}$ and $\mathcal{D}_{ij}$ as the 
{\it generalized acceleration} term and the {\it generalized shear} term, respectively. 

It is straightforward to see that $\mathcal{D}_{ij}$ reduces to the shear contribution 
in the the S-T limit. As $\tau_p \to 0$, we have $1/\tau_p \exp(\tau/\tau_p) \to \delta(\tau)$, 
and thus $\mathcal{D}_{ij}$ approaches the flow structure tensor $S_{ij}(r)$. 
As pointed out in \S1, it is exactly this flow structure tensor that is responsible for 
the shear contribution in eq. (1.1). 

In the S-T limit, $\mathcal{A}_{ij}$ can be evaluated as follows. For $\tau_p \to 0$, 
particle trajectories are close to those of tracers, so $B_{{\rm T}ij} \simeq B_{{\rm L}ij}$. 
For small time lag $\Delta \tau = \tau-\tau'$ (only small time lag is of interest here 
because of the exponential cutoffs in the integrand), $B_{{\rm L}ij}(\Delta \tau) \simeq 
(u'^2 -\frac{1}{2}a^2 \Delta \tau^2) \delta_{ij}$ 
where $u'$ and $a$ are, respectively, the rms flow velocity and the rms acceleration. 
Using this approximation for both $B^{(1)}_{{\rm T}ij}$ and $B^{(2)}_{{\rm T}ij}$ in eq. (2.13), we 
find $\mathcal{A}_{ij} = a^2 (\tau_{p1} -\tau_{p2})^2 \delta_{ij}$, which is exactly the same 
as the acceleration term in eq. (1.1). 


In \S 2.2 and \S2.3.3 we will show that, with our modeling of $\mathcal{A}_{ij}$ 
and $\mathcal{D}_{ij}$, eqs. (2.12), (2.13), and (2.14) also recover the large particle 
limit as well. 

Our formulation reflects the trajectory-crossing effect mentioned 
in \S 1. When setting $r$ to zero (or more exactly the particle size), 
the formulation is for two particles whose trajectories cross at 
time zero. Different from the model for the sling effect \cite[][]{fal02}, 
our model does not specify the physical mechanism how and when 
the trajectories of two particles cross. Instead the process leading 
to trajectory-crossing is considered indirectly from the backward 
separation behavior of the two particles.  From the following perspective, 
one may see a common feature shared by the sling effect and our 
model. The sling effect could be interpreted as a mechanism that 
contributes to the particle separation backward in time, and it gives a 
larger relative speed by increasing the contribution from the 
particles' memory of the flow velocity difference in the past. The latter is 
the point of our model. Therefore, we think that the sling effect can 
be accounted for in our model if the backward separation to be used 
in the model includes its contribution.

\subsection{Modeling $\mathcal{A}_{ij}$}
The formulation has been general so far. To proceed, we make assumptions for 
the trajectory correlation and structure tensors in eqs. (2.13) and (2.14). In this subsection, 
we evaluate the generalized acceleration term $\mathcal{A}_{ij}$.  

We use the usual assumption for $B_{{\rm T}ij}$ that the flow 
velocity viewed by a particle on its trajectory is the same as 
that by a tracer particle \cite[e.g.,][]{zai03a, zai03, zai06, aya08}, i.e., 
$B_{{\rm T}ij}= B_{{\rm L}ij}$. 
In a statistically stationary and isotropic flow, the Lagrangian 
correlation tensor only depends on the time lag, and can be written 
as $B_{{\rm L}ij} (\Delta \tau) = u'^2 \delta_{ij} \Phi (\Delta \tau)$ 
where $\Phi$ is the normalized temporal correlation function. 
We adopt the bi-exponential form for $\Phi$ \cite[see][]{saw91, zai03, zai06}, 
\begin{equation}
\begin{array}{lll}
\Phi(\Delta \tau; \tau_T, T_L)= {\displaystyle \frac{1}{2 \sqrt {1-2z^2} }} \Bigg[ \left(1 + \sqrt{1-2z^2}\right) \exp \left(
{\displaystyle -\frac{2 |\Delta \tau|}{ \left(1+ \sqrt{1-2z^2}\right) T_L } } \right)\\ 
\hspace{3.5cm} 
- \left(1-\sqrt{1-2z^2}\right) \exp \left ( { \displaystyle- \frac{2 |\Delta \tau|}{ \left(1 - \sqrt{1-2z^2} \right) T_L } } \right) \Bigg]
\end{array}
\end{equation}
where $T_L$ ($= \int \Phi (\Delta \tau) d \Delta \tau$) is the Lagrangian correlation timescale, 
and $z=\tau_T/T_L$ is the ratio of the Taylor micro timescale, $\tau_T$, to $T_L$. In addition to 
the time lag $\Delta \tau$, we have written $\Phi$ also as a function of $\tau_T$ and 
$T_L$  in eq. (2.16) for later convenience (see eq. (2.23)). 

We obtain $T_L$ using DNS results for the ratio of $T_L$ to the large-eddy 
turnover time, $T_E$.  $T_E$ is defined as the longitudinal integral scale, 
$L_1$, divided by the rms velocity, $u'$.  The length scale $L_1$  can be calculated 
from the relation, $\bar{\epsilon}=D u'^3/L_1$, using simulation results for 
the dimensionless coefficient $D$.  Defining a length scale $L$  as  $L = u'^3/\bar{\epsilon}$, 
we have $L_1= DL$. The large-eddy timescale is then given by $T_E = D u'^2/\bar{\epsilon}$ or 
$T_E = D T_e$ with $T_e$ defined as $T_e = u'^2/\bar{\epsilon} = L/u'$.   
\cite{yeu06a} found that the ratio of $T_L$ to $T_E$ is $\simeq 0.75$, and is 
essentially independent of the Taylor Reynolds number, $Re_\lambda$ 
(although this ratio may depend on the forcing and thus could be flow-dependent). 
Therefore we have $T_L = 0.75 D u'^2/\bar{\epsilon}$ or  $T_L = 0.75 D T_e$. 

Numerical simulations have shown that the coefficient, $D$, is 
Reynolds number dependent for $Re_\lambda \lsim 100$,  
but approaches a constant $\sim$ 0.4 for $Re_\lambda$ 
larger than several hundred \cite [e.g.,][]{yeu06b, ish09}. 
Therefore in the limit of large $Re_\lambda$, we expect 
$T_E = 0.3 u'^2/\bar{\epsilon}$. To also account for the $Re_\lambda$ 
dependence of $D$ at small $Re_\lambda$, we use $D= 0.4 (1+30 /Re_\lambda)$, 
which is obtained from fitting the numerical results in \cite{yeu06b}. We then have,   
\begin{equation}
T_L = 0.3 (1 + 30/Re_\lambda) u'^2/\bar{\epsilon}
\end{equation}
which is very close to that adopted by \cite{zai03, zai06} and the empirical 
formula given in \cite{saw08}. The definition of $Re_\lambda$ gives 
$u'^2 = Re_\lambda/\sqrt{15} u_{\eta}^2$, which will be used for normalization 
in our calculations.  

The Taylor micro timescale is defined as $\tau_T = (2 u'^2/a^2)^{1/2}$. 
The asymptotic behavior of the normalized acceleration variance, 
$a_0 = a^2/ (\bar\epsilon^{3/2}\nu^{-1/2}) $, at 
large $Re_\lambda$ has not been resolved by current simulations. 
Although $a_0$ is predicted to be constant at large $Re_\lambda$ 
limit by the Kolmogorov 41 theory \cite[see, e.g.,][]{vot98, zai03}, the 
intermittency corrections to the K41 theory may give it a power-law 
dependence on $Re_\lambda$ \cite[e.g.,][]{bor93}. 
For example, assuming that the temporal statistics of the dissipation rate along Lagrangian 
trajectories are the same as its spatial statistics in the Eulerian frame 
(which follows from the ergodic hypothesis and incompressibility; 
see \cite{bor93}), and using the intermittency theory by \cite{she94} for the 
dissipation rate statistics, we find that $a_0 \propto Re_\lambda^{0.133}$ 
(a similar result was obtained by \cite{bor93} 
using the log-normal intermittency model). This result is in impressive 
agreement with one of two formulas that well fit the results from 
simulations with resolution up to 2048$^3$ in \cite{yeu06b},                
\begin{equation}
a_0 = 1.9 Re_\lambda^{0.135} (1 + 85/Re_\lambda^{1.135})  
\end{equation} 
which goes like $Re_\lambda^{0.135}$ at large $Re_\lambda$. 
Therefore, one may expect that $a_0 = 1.9 Re_\lambda^{0.135}$ 
for asymptotically large $Re_\lambda$. However, 
the confirmation of this asymptotic behavior needs future simulations with higher 
resolutions \cite[][]{yeu06b} or more accurate experimental measurements \cite[][]{vot98}. We will use eq. (2.18) in our 
calculations. 
      
Eq. (2.16) approaches $\exp(-|\Delta \tau|/T_L)$ for $z \ll 1$ and 
$|\Delta \tau| \gg \tau_T$. The biexponential form is expected to be better than 
the single exponential form, $\exp(-|\Delta \tau|/T_L)$, because the Lagrangian 
correlation is believed to be smooth at small time lag, $|\Delta \tau| \to 0$. 
It can be easily shown that $\Phi \sim 1- \Delta \tau^2/\tau_T^2 $ 
for $|\Delta \tau| \ll \tau_T$, and thus satisfies the smoothness requirement.  

We insert eq. (2.16) into eq. (2.13) to calculate $\mathcal{A}_{ij}$. 
A lengthy but straightforward integration gives, 
\begin{equation}
\mathcal{A}_{ij} = u'^2 \delta_{ij} \frac{(\Omega_2 -\Omega_1)^2 \left(\Omega_1 \Omega_2 + (\Omega_1 + \Omega_2) {\displaystyle \frac{z^2}{2} } \right)  }
{(\Omega_1 + \Omega_2) \left(\Omega_1 + \Omega_1^2 + {\displaystyle \frac{z^2}{2}} \right) \left(\Omega_2 + \Omega_2^2 + {\displaystyle \frac{z^2}{2}}\right) } 
\end{equation}
where $\Omega_1 $ and $\Omega_2$ are defined as $\Omega_1 =\tau_{p1}/T_L$ and $\Omega_2 =\tau_{p2}/T_L$. Equation 
(2.19) correctly reproduces the acceleration term in the S-T limit. 
When $\Omega_1$, $\Omega_2 \ll z^2/2$, i.e., $\tau_{p_1}$, $\tau_{p2} \ll \tau_T^2/(2 T_L)$ 
(which is $ \sim \tau_{\eta}$ from the K41 phenomenology) , we have $\mathcal{A}_{ij} \to u^2 \delta_{ij} (\Omega_{2}-\Omega_{1})^2/ (z^2/2) = (\tau_{p2} -\tau_{p1})^2 a^2 \delta_{ij}$  where the definitions of $\Omega$, $z$ and $\tau_T$ have been used in the last step. 
As expected, for identical particles, the $\mathcal{A}_{ij}$ term is zero.

Results closely related or directly comparable to eq. (2.19) have been derived 
from several other models. \cite{wil83} considered the relative 
velocities in two limits with $\Omega \ll 1$ and $\Omega \gg 1$, 
and then gave a ``universal" solution by interpolating those two 
limits. For the small particle limit, they argued that the particle 
separation (back in time) can be neglected in the calculation of 
the particle velocity correlation as long as one of the two particles is very small. 
With this assumption, they find that the relative velocity 
variance, $\langle w_r^2 \rangle$, is given by (their eq. 19), 
\begin{equation}
\langle w_r^2 \rangle = u'^2 \frac{ (\Omega_2-\Omega_1)^2 } { (\Omega_1 +\Omega_2)(1+\Omega_1)(1+\Omega_2) }. 
\end{equation}
This result corresponds to our result for $\mathcal{A}_{ij}$ because 
neglecting the backward separation is essentially the same as 
neglecting the $\mathcal{D}_{ij}$ term. The latter can be justified if one of 
the two particles, say particle (1), has a very small friction time, i.e., 
$\tau_{p1} \to 0$. In that case, eq. (2.14) is approximately given by 
$\int_{-\infty}^{0} S_{{\rm T}ij} ({\bs r}; 0, \tau')\exp(\tau'/\tau_{p2}) 
d\tau'/\tau_{p2}$. For ${\bs r} \to 0$ as considered by \cite{wil83}, 
we have $S_{{\rm T}ij} ({\bs r}, 0, \tau') \to 0$ and thus $\mathcal{D}_{ij} \to 0$. 

It is obvious that, if $z$ is set to zero, eq. (2.19) for 
$\mathcal{A}_{ij}$ reduces to eq. (2.20). This is expected 
because, with $z=0$, $\Phi$ takes the same single-exponential form 
used in the derivation of eq. (2.20) by \cite{wil83}. 
Without the $z$ terms, eq. (2.20) does not reproduce the 
acceleration term in the S-T limit. \cite{kru97} generalized 
the model by \cite{wil83} and used a temporal velocity spectrum that 
accounts for the acceleration field in the flow (corresponding to 
a correlation function similar to our eq. (2.16) ) and obtained a formula that 
gives the acceleration term in eq. (1.1) in the S-T limit. 


In \S3.2, we find that for two particles of very different size 
the contribution to the relative speed from $\mathcal{A}_{ij}$ dominates 
over that from $\mathcal{D}_{ij}$, and thus  eqs. (2.19) or (2.20) 
can be used to estimate  the relative velocities between very different 
particles (although the justification above for neglecting $\mathcal{D}_{ij}$ 
is only for the case with at least one tiny particle).  
However, for similar particles, $\mathcal{D}_{ij}$ is 
the dominant term,  
and thus the result for $\Omega \ll 1$ by Williams and Crane 
is not valid for similar-size particles.
In the limit of particles with $\Omega \gg 1$, \cite{wil83}
considered particle separation. We will give 
more comments on their model in \S2.2.3 and \S 4.     

\cite{yuu84} derived a formula for the relative velocity 
with a shear term and an acceleration term. The derivation 
included an added mass term, $b\partial_t {\bs u}$, 
in the particle momentum equation (eq. (2.1)). The coefficient $b$, 
given by $3\rho_f/(2\rho_p +\rho_f)$, is small for solid particles 
in a gaseous flow and the added mass effect (Kruis and Kuster 1997)  
is not important. In that case the acceleration term given by \cite{yuu84} 
is exactly the same as eq. (2.20). This means that the acceleration term by 
Yuu is generally in agreement with our eq. (2.19) for 
$\tau_p \gg \tau_T^2/T_L \sim \tau_\eta$. His shear term thus 
corresponds to our $\mathcal{D}_{ij}$ term, which is the dominant term 
for similar particles. 
The problem in Yuu's shear term is that its derivation did not 
keep track of particle distance in the past, and thus the resulting 
shear term does not account for particle's memory of the flow velocity 
difference. This would underestimate the relative velocity 
between similar particles since the flow velocity 
difference was larger at earlier times when the particle separation was larger.     

With the Lagrangian correlation function, eq. (2.16), 
the first and third terms in eq. (2.13), representing the velocity 
variances of particle (1) and (2), $\left( v'^{(1)} \right)^2$ and $\left( v'^{(2)}\right)^2$, 
are given by, 
\begin{equation} 
\left(v'^{(1)} \right)^2 = u'^2 \frac{\Omega_1 + z^2/2}{\Omega_1 +\Omega_1^2 +z^2/2}{\rm,}\hspace{2mm} {\rm and} \hspace{2mm}  
\left( v'^{(2)}\right)^2 = u'^2 \frac{\Omega_2 +z^2/2}{\Omega_2 +\Omega_2^2 +z^2/2}. 
\end{equation}
For $\Omega_1$, $\Omega_2 \ll 1$, the particle rms velocity is close 
to the flow rms velocity $u'$. As expected, in the limit $\Omega_1$, $\Omega_2 \gg 1$, 
the sum of the two terms reproduces eqs. (1.2) and (1.3) for the relative velocity of large particles. 
This means that, in order to build a model that gives correct prediction 
in the limit of very large $\tau_p$, one needs to guarantee that the sum of 
the other two terms in the model,  i.e.,  the 2nd term in eq. (2.13) and $\mathcal{D}_{ij}$ 
(both from the cross correlation of particle velocities),  approaches zero 
as $\tau_p \to \infty$.  More specifically,  the sum of those two terms  
has to approach zero faster than $1/\tau_p $  so that they do 
not dominate over $\left(v'^{(1)} \right)^2$ and $\left(v'^{(2)} \right)^2$ 
given by eq. (2.21). We will show that the model presented in the next subsection 
for the $\mathcal{D}_{ij}$ term does satisfy this constraint, and hence 
eqs. (1.2) and (1.3) are correctly reproduced in our model. 

Finally,  we point out that the assumption we adopted that the 
temporal correlation of the flow velocity on an inertial particle's 
trajectory can be approximated by the Lagrangian correlation 
function may be invalid for large $\tau_p$.  The trajectory of 
a large particle can be very different from that of a tracer particle.  
For example, if $\tau_p \gsim T_L$ or $T_E$, the particle may 
not move significantly as the flow ``sweeps" by. Thus the correlation 
of the flow velocity along the trajectory of such a heavy particle 
may be better approximated by the Eulerian temporal correlation.  This means 
that the flow velocity correlation on a particle's trajectory could
make a transition from Lagrangian-like to Eulerian-like 
as $\tau_p$ increases.  In that case, replacing the Lagrangian 
correlation timescale  $T_L$  in eq. (2.21) by the Eulerian 
correlation timescale, $T_{Eu}$, would give a better estimate 
for the rms velocity of large particles.  $T_L$ in eq. (1.2) 
from  \cite{abr75}  should also be replaced by $T_{Eu}$.
If the Eulerian correlation timescale is larger than $T_L$ \cite[e.g.,][] {yeu89, kan91},  
our model would underestimate the relative speed by a factor of $(T_{Eu}/T_L)^{1/2}$ 
in the limit of $\tau_p \gg  T_L$.  A numerical study of the flow velocity correlation on particles' trajectory 
as a function of $\tau_p$ would be useful to improve our model. 
  
\subsection{Modeling $\mathcal{D}_{ij}$ }        
In order to evaluate the generalized shear term, $\mathcal{D}_{ij}$, 
we need to model the trajectory structure tensor, $S_{{\rm T}ij}$,  
which has not been directly measured.  
The tensor can be formally written as 
\begin{equation} 
S_{{\rm T}ij} =  \int \int \left \langle \delta u_{i}^{(p)}(\tau) \delta u_{j}^{(p)}(\tau')| \bs{\rho}, \bs{\rho}'  \right \rangle 
P(\bs{\rho}, \bs{\rho'}; \bs{r}, \tau, \tau') d\bs{\rho}  d\bs{\rho'}
\end{equation} 
where $P$ is the joint probability distribution of the particle separations, 
$\bs{\rho}$ and  $ \bs{\rho'}$, at $\tau$ and $\tau'$,  respectively,  
and we have used $\delta u_{i}^{(p)}(\tau)$ ($\equiv u_{i} (\bf{X}^{(1)}(\tau)) - u_{i}(\bf{X}^{(2)}(\tau)) $)  
and $\delta u_{j}^{(p)}(\tau')$ to denote the flow velocity difference 
``seen" by the two particles. The ensemble average term in the 
integrand is the velocity difference correlation conditioned on the 
particle separations.  A series of assumptions need to be made 
for the estimate of $S_{{\rm T}ij}$ since both the conditional 
correlation and the joint probability of the particle separations 
are unknown.  

The conditional correlation depends on both the separations, 
$\bs{\rho}$ and $\bs{\rho'}$, and the two times, $\tau$ and 
$\tau'$. It may also have a direct dependence on $\bs{r}$, 
in addition to that through $\bs{\rho}$ and $\bs{\rho'}$. This possible 
dependence is neglected here. We assume that the 
amplitude of the velocity difference $\delta u_{i}^{(p)}(\tau)$ on 
particles' trajectories can be approximated by the Eulerian 
velocity difference, $\delta u_{i} (\bs{\rho}, \tau)$, across 
the separation $\bs{\rho}$. This assumption is similar to the 
independence hypothesis by \cite{cor59}  (note the 
similarity between eq. (5) in \cite{cor59} and our 
eq. (2.22); see also \cite{shl74} ). Assuming that the displacement of a fluid particle 
(from its initial position) is statistically independent 
from the particle's current velocity,  \cite{cor59} gave 
a relation between the Lagrangian correlation function 
and the Eulerian correlation. The hypothesis by Corrsin 
essentially neglects a constraint between 
the particle displacement and the 
velocity along the Lagrangian trajectory 
(i.e., the latter equals the time derivative of the former).    
In the case of inertial particles, the constraint neglected in 
our approximation is $d^2\bs{\rho}/dt^2 = ( \delta \bs{u}^{(p)}  -  d\bs{\rho}/dt )/ \tau_p$.   
The existence of this constraint could give rise to a 
statistical correlation between $\delta \bs{u}^{(p)}$ and $\rho$. 
For example, a larger particle separation at a given time 
suggests a larger (on average) velocity difference in 
the past, and perhaps a higher possibility of a large 
velocity difference right at that time. Therefore, relative to 
the Eulerian velocity difference across a fixed separation, the 
velocity difference along the particles' trajectories may have a stronger dependence 
on the particle separation. 
Setting $\delta u_{i}^{(p)}$ to the Eulerian velocity difference, $\delta u_{i} (\bs{\rho})$, 
could thus underestimate the dependence of $\delta u_{i}^{(p)}$ on $\rho$.  
The uncertainty and reliability of our assumption here are subject to 
tests by future numerical experiments.  

From the assumption above, we have $\langle \delta u_{i}^{(p)}(\tau) \delta u_{j}^{(p)}
(\tau')| \bs{\rho}, \bs{\rho}' \rangle \simeq  \langle \delta u_{i} (\bs{\rho}, \tau) \delta u_{j} 
(\bs{\rho}', \tau') \rangle$. 
The correlation of the Eulerian velocity differences at two times 
is also unknown and further approximations are needed.    
We first assume that it can be written as a product of 
a separation dependence term and a time-lag dependence term. 
The separation dependence term is then assumed to take the 
form of the Eulerian structure function of the flow, $S_{ij}$. 
Note that, the conditional correlation actually depends on 
two separations, $\bs{\rho}$ and $\bs{\rho}'$, but for 
simplicity we will approximate it by $S_{ij}({\bs R})$, at a single 
separation, ${\bs R}$, characteristic of the particle distances 
between $\tau$ and $\tau'$.  The choice for the separation $\bs{R}$ 
as a function of $\bs{\rho}$ and $\bs{\rho}'$ will be discussed later.  
As a function of the two stochastic vectors, $\bs{R}$ is also stochastic. 
 
The time-lag dependence accounts for the temporal correlation 
between the flow velocity differences ``seen" by the two particles.  
This correlation depends on the persistence of the ``structure" 
in question, which is a function of the ``structure size". Here 
the size is essentially the distance between the two particles. 
Associated with each structure, there is a correlation timescale, 
$T_R$. 
To estimate $T_R$, one needs to pick up a size to characterize 
the structure that corresponds to particle distances at time 
between $\tau$ and $\tau'$. We will take the size to be the 
same as the particle separation $R$ to characterize the 
separation dependence. We point out that there is no 
physical motivation for this particular choice, and that it could 
be better to choose the distance, $R_m$, at the earlier one of 
the two times $\tau$ and $\tau'$, $\min(\tau, \tau')$, assuming the 
persistency of a structure is determined by its initial size. The difference in the results from 
the two choices will be discussed in \S3. The time lag dependence 
is assumed to take the same function form as eq. (2.16) for 
the Lagrangian temporal correlation in \S2.2.   
Namely, we set the time-lag dependence to be $\Phi(\tau-\tau'; \tau_{TR}(R), T_R(R) )$,  
where the Lagrangian timescale, $T_L$, in eq. (2.16) has been 
replaced by $T_R(R)$ and $\tau_T$  by $\tau_{TR} (R)$. 
The timescale $\tau_{TR} (R)$ is an analogue of the Taylor 
micro timescale for a structure of size $R$ \cite[see][]{zai03, zai06}.    


The conditional correlation is now approximated by 
$S_{ij} ({\bs R})\Phi \big(\tau -\tau'; \tau_{TR}(R), T_R(R) \big)$. 
In order to estimate the trajectory structure function, in principle 
one needs to integrate the conditional correlation over the 
distribution $P(\bs{\rho}, \bs{\rho'})$ (eq. 2.22), or equivalently 
over the distribution, $P(\bs{R}) $, of $\bs{R}$ (in our 
approximation the conditional correlation depends on the 
separations $\bs{\rho}$ and $\bs{\rho}$ only through $\bs{R}$).  
We take a simple approach here.  We set the integral to be 
equal to the conditional correlation at a distance corresponding 
to the rms of $\bs{R}$.   
This means that the particle distance at a given time is assumed to 
be single-valued, i.e., the distribution $P(R)$ is a delta function 
at the rms of $\bs(R)$.  A rough justification of this approximation will be 
given in \S2.3.3. The direction of $\bs{R}$ probably has a random 
distribution due to the turbulent dispersion,  and we will average 
the structure tensor $S_{ij} (\bs {R})$, 
over the direction distribution of $\bs{R}$.   

For simplicity in notations, hereafter we will use $R$ to denote the rms 
length of $\bs{R}$, i.e., $R^2 = \langle {\bs{R}^2} \rangle$ (note that this 
is different from the conventional notation that $R$ denotes the 
length of a vector $\bs{R}$, i.e., $R^2 = \bs{R}^2$, without ensemble 
averaging). Similarly,  $\rho(\tau)$ will denote the rms length of 
$\bs{\rho}(\tau)$. 
We refer to $R$ and $\rho$ as ``rms distance" or simply ``distance", 
while using ``separation" for the corresponding stochastic vectors.

The approximations for the trajectory structure tensor are now complete,      
\begin{equation} 
S_{{\rm T}ij} ({\bs r}; \tau, \tau') \simeq \big\langle S_{ij} ({\bs R})\big\rangle_{\rm ang} 
\Phi \big(\tau -\tau'; \tau_{TR}(R), T_R(R) \big) 
\end{equation}
where the ensemble average for $S_{ij}({\bs R})$ is over the 
direction distribution of $\bs{R}$.  The angular average will be carried out in \S2.3.3.  


Finally we need to specify the rms distance $R$ as a function of 
$\rho(\tau)$ and $\rho(\tau')$. 
If the flow velocity difference scales with the distance as a power 
law, which is probably the case in well-developed, homogeneous 
and isotropic turbulence, a good choice would be 
\begin{equation}
R (\tau, \tau')= \big(\rho(\tau) \rho(\tau')\big)^{1/2}.   
\end{equation}
We note that, in their assumption for the Lagrangian structure tensor, \cite{zai03a} 
apparently set $R$ to be the particle separation at the earlier 
time of $\tau$ and $\tau'$. We argue that eq. (2.24) is probably 
a better assumption because $S_{{\rm T}ij}$ is expected to 
be zero if either $\rho(\tau)$ or $\rho(\tau')$ is zero.   

The assumption, eq. (2.24), is expected to be valid when 
$\rho$ and $\rho'$ are in the same length scale subrange 
since in that case the scaling of velocity difference across 
the two separations follows the same power-law (see, \S 2.3.1). 
On the other hand, if $\rho$ and $\rho'$ are in different 
subranges,  the structure function across $R$ defined 
by eq. (2.24) may not correctly represent the product 
of the velocity difference amplitudes across $\rho$ 
and $\rho'$. Fortunately, we find that this does not 
significantly affect our prediction of the relative speed  
based on a consideration of the $\Phi$ term in the trajectory structure 
tensor (the temporal correlation of the velocity differences). 
The $\Phi$ term gives an exponential cutoff when the time 
lag is large, i.e., for very different  $\tau$ and $\tau'$. 
Note that, if $\rho$ and $\rho'$ are in different subranges,  
in general $\tau$ and $\tau'$ would  also be very different 
(although it is possible that the particle pair experiences 
a large separation change during a short time interval, these 
extreme events should be rare, and not affect the low order 
statistics, i.e., the second order particle structure function, 
we study here). Therefore, the exponential cutoff from the 
$\Phi$ term would suppress the contribution from very different $\rho$ 
and $\rho'$ to the integral for $\mathcal{D}_{ij}$. 
In other words, the temporal decorrelation of the velocity difference 
over large time lags suggests that the main contribution to $\mathcal{D}_{ij}$  
is probably from similar $\rho$ and $\rho'$, where eq. (2.24) 
is valid. 





\subsubsection{The Flow Structure Tensor and the Timescales}
In homogeneous and isotropic turbulence, the Eulerian structure 
tensor at a separation $\bs{l}$ can be written as 
\cite[e.g.,][]{mon75},  
\begin{equation}
S_{ij}({\bs l}) = S_{nn} (l) \delta_{ij} + \big( S_{ll} (l) - S_{nn}(l) \big) \frac {l_i l_j} {l^2} 
\end{equation}
where $S_{nn}$ and $S_{ll}$ are, respectively, transverse and longitudinal structure functions. 
For an incompressible velocity field, they are related by,  
\begin{equation}
S_{nn}(l) =S_{ll} (l) + \frac{l}{2} \frac{d S_{ll}(l) }{dl}.    
\end{equation} 
The longitudinal structure function, $S_{ll}$, in different 
ranges of length scales is given as follows.  

As mentioned in \S1, in the viscous subrange, 
$S_{ll}$ is given by,   
\begin{equation} 
S_{ll} (l) =\frac {\bar{\epsilon}}{15 \nu} l^2 {\rm, \hspace{2mm} for} \hspace{2mm} l \lsim \eta. 
\end{equation} 

In the inertial subrange, we have, 
\begin{equation} 
S_{ll} (l) = C (\bar{\epsilon} l)^{2/3} {\rm,  \hspace{2mm} for } \hspace{2mm} \eta \lsim l \lsim L_1
\end{equation} 
where the coefficient $C$ for the velocity scaling in the inertial 
range is believed to be universal and will be set to $C=2$ 
\cite[][]{mon75, zai03a}. 

For $l$ larger than the integral scale, $S_{ll}$ is constant, 
\begin{equation} 
S_{ll} (l) = 2 u'^2 {\rm,  \hspace{2mm} for } \hspace{2mm} l \gsim L_1.   
\end{equation}

The characteristic scale at which $S_{ll}$ switches from the viscous-range 
scaling to the inertial-range scaling can be obtained by equating 
eqs. (2.27) and (2.28). This gives a transition scale of $(15 C)^{3/4} \eta$, 
which is about $13 \eta$ for $C=2$.  It is consistent with the simulation 
results given in \cite{ish09}, where the switch occurs at about a few tens 
of Kolmogorov scale, between $\eta$ and the Taylor micro scale $\lambda$. 
The scaling changes from eq. (2.28) to eq. (2.29) at $l \simeq (2/C)^{3/2} L $ 
where $L = u'^3/\bar{\epsilon}$, 
as defined earlier.  
This is also in general agreement with Fig. 7b in \cite{ish09}, which shows 
that $S_{ll}$ becomes constant at $\simeq 0.5-3 L_1$ ($L_1 \simeq 0.4 L $).    
   
We will use the following formula from \cite{zai06} to connect the velocity scalings 
in different subranges,
\begin{equation}
S_{ll} (l) = 2 u'^2 \left[ 1- \exp \left( - \frac{l}{(15C)^{3/4} \eta} \right) \right]^{4/3}  
\left (\frac{l^4}{l^4 + (2/C)^6 L^4} \right)^{1/6}.          
\end{equation} 
The transverse structure function $S_{nn}$ then follows from eq. (2.26). 

The timescale $T_R$ as a function of the separation in eq. (2.23) 
has different scalings in the three subranges above as well. 
The theoretical model by \cite{lun81} found that in the 
viscous range, 
\begin{equation} 
T_R(l) = \sqrt{5} \tau_{\eta} {\rm, \hspace{2mm} for} \hspace{2mm} l \lsim \eta 
\end{equation}
which was later confirmed by numerical simulations \cite[][]{gir90}.

In the inertial range, the 
similarity argument suggests that,  
\begin{equation} 
T_R(l) = C_2 \bar{\epsilon}\hspace{0.8mm}^{-1/3} l^{2/3} {\rm,  \hspace{2mm} for } \hspace{2mm} \eta \lsim l \lsim L_1.   
\end{equation}   
Following \cite{zai06}, we will take the coefficient 
$C_2 =0.3$ in our calculations. Eqs. (2.31) and (2.32) 
connect at $\simeq (\sqrt{5}/C_2)^{3/2} \eta$, which is 
$\simeq 20 \eta$ for $C_2=0.3$. This is a little larger 
than the corresponding transition scale ($13 \eta$) for $S_{ll}$.

For $l \gg L_1$, the flow velocities across $l$ 
are independent, $T_R$ is thus expected to 
be the correlation timescale of the velocity 
along the trajectory of each particle, 
which is approximately given by the 
Lagrangian correlation timescale, $T_L$ 
\cite[see][]{zai03a},   
\begin{equation}
T_R(l) = T_L {\rm ,  \hspace{2mm} for } \hspace{2mm} l \gsim L_1.  
\end{equation}
This connects to the inertial-range scaling at about 
$\bar{\epsilon}^{1/2} (T_L/C_2)^{3/2}$, which is $\simeq L$ 
using $C_2 =0.3$ and $T_L \simeq 0.3 u'^2/\bar{\epsilon}$.  
  
Similar to the case for $S_{ll}$, a formula is used to 
connect $T_R$ in different subranges, 
\begin{equation} 
T_R(l) = T_L \left [1 -\exp\left(-\left(\frac{C_2}{\sqrt{5}}\right)^{3/2} \frac{l}{\eta }\right) \right]^{-2/3}\left(\frac{l^4}{l^4 + T_L^6 (\bar \epsilon)^2/C_2^6 } 
\right)^{1/6}  
\end{equation} 
which is again adopted from \cite{zai06}. 

By analogy to the definition of the Taylor micro timescale, $\tau_T$,  we 
estimate $\tau_{TR}(l)$ by $\tau_{TR}^2  \simeq  2  \langle \delta u (l)^2 \rangle /  \langle \delta a(l)^2 \rangle$ 
where $\delta u (l) $ and $ \delta a (l)$ are the velocity and acceleration 
difference across a distance $l$. This formula is simply a generalization of 
$\tau_T$ for the one-particle Lagrangian correlation to that  for two-particle 
Lagrangian structure function. The correlation length scale of the 
acceleration field is expected to be short, probably $\sim \eta$.  In that case, 
for $l$  in the inertial range $\langle \delta a(l)^2 \rangle \simeq 2 a^2$ with $a$ the 
rms acceleration and thus $\tau_{TR} \propto \delta u(l) \sim  l^{1/3}$.  From $T_R(l) \sim l^{2/3}$, 
we have $\tau_{TR}(l) \propto T_R(l)^{1/2}$.  
In our calculations, we will use 
$\tau_{TR}= \tau_T  (T_R /T_L)^{1/2}$. Note that this formula gives 
$\tau_{TR} = \tau_T$ for $l \gsim L$, as expected. It gives a constant $\tau_{TR} $ for $l \to 0$.  
 \cite{zai03, zai06}  assumed that $\tau_{TR} (l) = (\tau_T/T_L) T_R(l) = z T_R(l)$ without 
providing a physical motivation.  Our calculations find that the two different 
assumptions for $\tau_{TR}$ do not give significant difference in the predicted 
relative speed. 
  
\subsubsection{Particle Pair Dispersion}
We now consider the rms distance, $\rho (\tau)$, of two particles as 
a function of time $\tau$, which is needed to evaluate $R$ (eq. (2.24)).  
As mentioned earlier, the specific question we ask here is how particles 
separate from each other backward in time, given their separation, $\bs{r}$, 
at time zero (we will also refer to $r$ as the initial distance for the backward 
dispersion from the viewpoint of the reversed time direction, although 
with normal time direction it is the final distance of the two particles 
in question). 

The study of turbulent dispersion of inertial particles started   
only recently \cite[][]{bec07, fou08, bec09}, 
\cite{bec09} gave a detailed report of simulation results for 
the separation behavior forward in time. They found that there are 
two temporal regimes with different separation behaviors:  
a transient regime representing the relaxation of the particle 
velocity toward  the flow velocity,  and a later regime where the 
particle pairs separate in a similar way as tracer particles. 
The transient regime lasts for about a friction timescale. For $St \gsim 3$, 
a ballistic separation is found in the transient regime and  
the separation speed is equal to the initial velocity difference.  
The ballistic separation is due to the particles' memory for 
a period of $\sim \tau_p$. 
In the later phase, the particle separation is found to 
follow the Richardson-Obukhov separation law.  

Although the study by \cite{bec09} is for dispersion 
forward in time, their results provide a very useful guideline 
for us, because we are are not aware of any investigations for 
the backward dispersion of inertial particles. 
A separation behavior similar to that found by \cite{bec09} 
will be used in our calculations (\S 3.1.3). We will show 
that a combination of an earlier ballistic phase and a 
later tracer-like phase gives quite good fit to the numerical simulation 
results for the relative speed by \cite{wan00}.  For $St \lsim 1$, the particle separation appears to 
increase slower than linearly with time in the 
transient phase according to Fig. 5 in \cite{bec09}.
However,  no function fit to the separation behavior 
in this regime is given by \cite{bec09}.  For simplicity, 
we will assume that the separation is ballistic for the 
early phase of all particles. This 
assumption gives rise to uncertainty in our prediction 
for the relative speed for particles with $St \lsim 1$.
  
In order to understand the results of our model with 
the two-phase separation, we need to know the 
effect of each phase, and thus we first consider two 
simplified cases assuming a complete ballistic 
behavior (\S3.1.1) and a complete tracer-like behavior (\S3.1.2), 
respectively. The simplified cases give a very useful illustration
for the effect of particle separation on the predicted 
relative speed.  For example, for both ballistic and tracer-like separation behaviors, 
the relative speed as a function of the Stokes number can be 
explained from an approximate analysis of the integral 
equation for $\mathcal {D}_{ij}$. In particular, the analysis 
gives physical insights on the scaling of the relative speed 
with the Stokes number in the inertial range for the 
monodisperse case. Once the two simplified cases are understood, 
it is straightforward to interpret the prediction of our model with the 
more realistic separation behavior by a combination 
of two phases (\S3.1.3).   

For ballistic motions, the particle distance goes linearly with time. 
Given the particle distance, $r$, at time zero, the separation as 
a function of $\tau$ is,       
\begin{equation}
\rho^2(\tau) = r^2 + \langle w^2 \rangle \tau^2  
\end{equation}   
where $\langle w^2 \rangle =\langle ( {\bs v}^{(1)} -{\bs v}^{(2)} )^2 \rangle =S_{{\rm p} ii}(\bs{r})$ 
is the 3D relative velocity variance of the two particles at time zero.  
The separation speed at any time is taken to be the same as that 
at time zero. Recall that in our notation $\rho(\tau)$ is the rms 
of $\bs{\rho}(\tau)$. The relative velocity variance, $\langle w^2 \rangle$, 
in eq. (2.35) is unknown and is directly related to the radial relative 
speed under pursuit. We will build an implicit equation for 
$\langle w^2 \rangle$  and $\langle w_r^2 \rangle$ in \S 3, which is then 
solved self-consistently.

The exponential cutoffs in the integrand of eq. (2.14) imply that the 
primary contribution to ${\mathcal D}_{ij}$ is from  $-\tau_{p1} 
\lsim \tau \le 0$ and $-\tau_{p2} \lsim \tau' \le 0 $. This means that if the 
ballistic behavior lasts for about $\tau_p$ \cite[][]{bec09}, 
then using eq. (2.35) at all times when integrating eq. (2.14) 
may give an acceptable order-of-magnitude estimate for 
${\mathcal D}_{ij}$, even though the separation is not ballistic at later time. 
 
\cite{wil83} \cite[and also][]{kru97} take the particle motions to 
be ballistic in their calculations for the limit with large particles 
($\tau_p \gg T_L$). Assuming the velocity correlation between 
two particles can be neglected for the purpose of estimating the 
particle separation, they set the linear separation 
rate to $\left( (v'^{(1)})^2 +(v'^{(2)})^2 \right)^{1/2}$. We note that this 
assumed separation rate may be a good approximation only 
for very large particles. In the case of small to intermediate particles, 
the velocities of nearby particles are correlated 
and the separation rate is smaller than given by \cite{wil83}. Using 
their separation speed for those particles would overestimate 
the relative speed because both $S_{ij}$ and $\Phi$ in eq. 
(2.23) increase with the particle distance, as can be seen from 
eqs. (2.30) and (2.34).    

In the second simplified case, we will consider the separation 
behavior similar to that of tracer particle pairs \cite[see]
[for detailed reviews on the pair dispersion of tracers] 
{fal01, sal09}. We are particularly interested in the effect of
Richardson-Obukhov separation (the R-O separation hereafter) 
phase  found in  the forward dispersion of inertial particle pairs.    
The R-O separation law is written as  
\begin{equation}
\rho^2(\tau) \propto g \bar{\epsilon} |\tau|^3  
\end{equation} 
where the dimensionless coefficient $g$ is known as 
the Richardson constant. \cite{bec09} did not 
give best-fit values for $g$ in the late-phase separation.  
Apparently $g$ needs to be adjusted to fit the simulation 
results (\cite{bec09}), and it probably has a Stokes 
number dependence.  Since the value of $g$ for inertial particles is unknown,  
we will first use $g$ measured for tracer pairs as a reference. 
For tracers, theoretical models and direct numerical 
simulations by \cite{saw05} show that the dispersion 
backward in time is significantly faster than the forward 
dispersion. Experimental measurements by \cite{ber06} 
found that $g =0.55$ for the forward dispersion, and $g=1.15$ 
for the dispersion backward in time. We will take $g=1.15$ as 
a reference value since it is the backward dispersion 
that is relevant in our problem. In \S 3.1.3, we show that 
a two-phase separation with $g \sim 1$ in our model 
gives good fit to the numerical results for the relative speed.   

When the particle separation becomes larger than the 
integral scale, the flow velocities ``seen" by the particle pair 
are uncorrelated and the separation is expected to be 
diffusive. Thus we will switch from the R-O law to the diffusive 
separation when the particle distance exceeds $L$, i.e., 
we set  $\rho^2(\tau) \simeq 2 D |\tau| $ for $ \rho \ge L$ 
where the coefficient $D$ is given by $D= 6 u'^2 T_L$. Note that $D$ 
here is for the 3D diffusion, and it is twice larger than $D$ 
for 1-particle diffusion (from the Taylor theorem) because the 
rms relative velocity of two faraway particles is $2 u'^2$. 
We find that  the exact separation behavior in the range 
$\rho \gsim L$ in the diffusive regime is not important for the 
integration of eq. (2.14) because $S_{ij}$ and $T_R$ in eq. 
(2.23) becomes constant in this range of 
the particle distance.



The smallest initial separation considered by \cite{bec09} is 
about $1\eta$. For an initial separation much below $\eta$,
there probably exists an initial exponential separation phase (similar to that of 
tracer pairs at small separations). 
The Lyapunov exponents for particles with $St \lsim 2$
have been computed from simulation by \cite{bec06}.  
Although the exponential regime would also exist for larger particles 
as implied by the chaoticity of the dynamics (see, e.g., \cite{bec07} for a 
theoretical model which predicts that the Lyapunov exponent 
decreases as $St^{-2/3}$ for large $St$), its relevance at finite 
(but below $\eta$) scale separation is questionable at least in the large $St$ limit.
Here, we will not consider the exponential phase, 
since it is unknown how long it lasts and how it 
connects with the later phases. In most of our calculations 
we will give results for $r \sim \eta$.  In the simulations by 
\cite{wan00} and \cite{zho01} that we will use to test our 
model, the relative speed is measured at a distance of $\eta$, 
thus it is sufficient to use Bec {\it et al.}'s result as a guideline in the 
comparison with those simulation results. It is straightforward to 
incorporate an exponential phase into our model, and     
once the detailed separation behavior of all particles 
at distances well below $\eta$ is known, 
our model can predict the relative speed at any separation.

Based on our physical picture, a careful consideration of the 
particle pair dispersion is necessary for an accurate estimate 
of the relative velocity. It is likely that the physics of turbulent 
dispersion of particle pairs is implicitly incorporated into the 
quite successful model by Zaichik and collaborators. In fact, 
their equation for the joint pdf of the particle separation and the 
relative velocity can be regarded as one for the pdf of the 
particle distance when integrated over the relative velocity (phase) 
space. For a comparison of our results with Zaichik {\it et al}.'s 
model, it is useful to see how tracer pairs separate in 
their framework, i.e., the prediction for particle separation 
from their formulation in the limit $\tau_p \to 0$. 
In this limit, the joint pdf equation can be reduced to an 
equation for the pdf of the particle distance.
The pdf equation turns out to be in the same form as that 
suggested by \cite{ric26}. With structure functions and 
timescales given in \S2.3 (same as in \cite{zai06}), 
the equation suggests that the Richardson constant, 
$g$, is about $3$, significantly larger than measured 
from experiments and numerical simulations. 
It is not clear whether and how the quasi-normal 
assumption made by Zaichik {\it et al}. to close the moment equations 
of the joint pdf equation may affect the particle dispersion, or what it 
physically corresponds to regarding the separation.

\subsubsection{Average over the Direction of $\bs{R}$}
We calculate the average of $S_{ij}$ over the direction 
of $\bs{R}$ in eq. (2.23). For $\bs {R}$ at any given time, 
we define a separation difference $\Delta {\bs R} = {\bs R} - {\bs r}$, 
the change of the separation from $\bs{r}$ at time zero. 
The direction of the separation difference is expected to be 
completely random if the flow velocity is statistically isotropic. 
This means that $\langle \Delta R_i \Delta R_j \rangle_{\rm ang} 
= \frac{\Delta R^2}{3} \delta_{ij}$.  
We set ${\bs l}$ in eq. (2.25) to be ${\bs R} = {\bs r} + \Delta \bs{R}$ 
and take the average over the direction of $\Delta {\bs R}$. 
A rigorous derivation of this average needs to consider the dependence of 
$S_{ij}({\bs R})$ on $\Delta {\bs R}$ 
through the length $R= |{\bs r}+ \Delta {\bs R} |$ (e.g., in $S_{ll}$ and $S_{nn}$) 
and that through the tensor $R_i R_j$ simultaneously. However, 
the derivation is very complicated and cannot be done analytically. 
For simplicity, we neglect the dependence through $R$  in the 
averaging process and only consider the average of $R_i R_j$ 
over the $\Delta {\bs R}$ direction. With this approximation, we find,  
\begin{equation}
\langle S_{ij}({\bs R}) \rangle_{\rm ang}= \left [\left (\frac{2}{3 }+ \frac{r^2}{3R^2} \right) S_{nn}(R) + \left (\frac{1}{3} - 
\frac{r^2}{3R^2} \right)  S_{ll} (R) \right] \delta_{ij}+ \big(S_{ll}(R) -S_{nn}(R)\big)
\frac{r_ir_j} {R^2}
\end{equation}
where we have used $\langle \Delta R_i r_j \rangle_{\rm ang} =0$ and $(\Delta R)^2 = R^2 -r^2 $.

The generalized shear term, ${\mathcal D}_{ij}$, then follows from eqs. (2.14), (2.23) and (2.37),   
\begin{equation}
\begin{array}{lll}
{\displaystyle {\mathcal D}_{ij}(r) = \int_{-\infty}^0 \frac {d\tau}{\tau_{p1}} \int_{-\infty}^0 \frac {d\tau'}{\tau_{p2}} 
\bigg\{ \left [\left (\frac{2}{3 } + \frac{r^2}{3R^2} \right) S_{nn}(R) + \left (\frac{1}{3} - 
\frac{r^2}{3R^2} \right)  S_{ll} (R) \right] \delta_{ij} }  \\ \hspace{3.5cm}
{\displaystyle + \big(S_{ll}(R) -S_{nn}(R)\big) \frac{r_ir_j} {R^2} \bigg\}   \times \Phi \big(\tau -\tau'; \tau_{TR}(R), T_R(R)\big)} \\\hspace{3.5cm} 
{\displaystyle \times \exp \left( \frac {\tau}{\tau_{p1}} \right)  
\exp \left( \frac {\tau'}{\tau_{p2}} \right)}. 
\end{array}
\end{equation} 
We will numerically integrate this equation in \S3 using the structure 
functions $S_{ll}$ and $S_{nn}$ and the timescales $\tau_{TR}$ and $T_R$ 
given in \S2.3.1, and particle pair separation laws for $\rho$ and 
$R$ given in \S2.3.2.  

In eq. (2.38), the structure functions, the timescale $T_R$,  and 
hence $\Phi$ (eq. (2.16)) increase with $R$, which increases 
with $|\tau|$ and $|\tau'|$. Together with the exponential cutoffs, this 
suggests that the integrand in eq. (2.38) peaks at $\tau \simeq -\tau_{p1}$ and $\tau' \simeq -\tau_{p2}$. 
Therefore, the main contribution to the integral is from 
$\tau \sim -\tau_{p1}$ and $\tau' \sim -\tau_{p2}$ 
(if $\tau_{p1}$ and $\tau_{p2}$ are not very different) 
and an important factor to determine the value of 
the integral is the distance, $R$, at $\tau \sim -\tau_{p1}$ 
and $\tau' \sim -\tau_{p2}$, which we will refer to as the 
primary distance. If $\tau_{p1}$ and $\tau_{p2}$ are both 
very small so that the primary distance is close to $r$, 
$\mathcal{D}_{ij}$ would approach $S_{ij}(r)$ as expected 
for the S-T limit. If the primary distance is much larger than 
the particle distance at time zero, $r$ , the $r^2/R^2$ 
terms in eq. (2.38) can be neglected and $\mathcal{D}_{ij} \propto \delta_{ij}$ 
(meaning that $\mathcal{D}_{ll}= \mathcal{D}_{nn}= \mathcal{D}_{ii}/3$).

If $\tau_{p1}$ and $\tau_{p2}$ are in the large limit and 
the primary distance is much larger than the length scale $L$, 
then typically the structure tensor in the integral in eq. (2.38) 
is  $\sim 2 u'^2 \delta_{ij}$ (see eq. (2.29)), and the $\Phi$ term 
would be the same as eq. (2.16) because $T_R =T_L$ (eq. (2.33)) 
for $R \gsim L$. This means that the integrand of eq. (2.38) would 
be the same as that in the 2nd term on the r.h.s. of eq. (2.13) 
(with $B_{{\rm T}ij}$ given by the assumptions in \S 2.2) 
for the range of  $\tau$ and $\tau'$ with $R(\tau,\tau') \gsim L$.  
On the other hand, at smaller values of  $|\tau|$ and $|\tau'|$ (with $R <L$),  
the integrand for $\mathcal{D}_{ij}$ is smaller than that in the latter.  
Thus $\mathcal{D}_{ij}$ given by (2.38) does not exactly cancel out the 
2nd term on the r.h.s. of eq. (2.13). We find both numerically 
and analytically that the sum of the two terms decreases faster than 
$1/\tau_p$ in the limit $\tau_p \to \infty$ (it goes like $\tau_p^{-3/2}$ for 
the assumption of the ballistic separation and like $\sim \tau_p^{-2}$ 
for the tracer-like separation behavior). Therefore,  in the limit of 
large friction timescales our model satisfies the constraint discussed 
at the end of \S2.1, namely,  a good model needs to give a particle velocity 
correlation that decreases with $\tau_p$ faster than the other terms.

As discussed earlier, we used the rms distance, $\rho$ and $R$,
in our estimate of the trajectory structure tensor, $S_{{\rm T}ij}$, 
while a rigorous derivation needs to consider the probability 
distribution function of the separation and take the average 
of $S_{{\rm T}ij}$ over this distribution. Here we give a 
justification for this approximation\footnotemark\footnotetext
{In the justification argument, $R$ is taken to be a stochastic 
variable, the length of the stochastic vector $\bs{R}$, although 
at all other places in the paper $R$ refers to the rms length 
of $\bs{R}$. The rms distance will be written explicitly as 
$\langle R^2 \rangle^{1/2}$ in the argument here. 
This change of notation is only for the present paragraph.}.
First, if the separation is in the viscous range, 
$R \lsim \eta$,  $S_{ij}$ has a quadratic dependence 
on the separation (eq. (2.27)) and $\Phi$ is independent of 
$R$ (eq. (2.31)). This suggests that the approximation 
is exact since the average of the quadratic dependence 
over the separation distribution is exactly the square 
of the rms distance. If $R$ is in the inertial range, then 
$S_{ij} \propto R^{2/3}$, and, roughly speaking, the $\Phi$ 
term provides another factor of $R^{2/3}$ (from the timescale $T_R(R)$, eq. (2.32), 
over which the flow velocity difference ``seen" by the particles is correlated). 
Therefore, using a single rms distance to replace the distance 
pdf would overestimate $S_{{\rm T} ij}$ (and hence $\mathcal{D}_{ij}$) 
by a factor of $\simeq \langle R^2 \rangle^{2/3}/\langle R^{4/3} \rangle$. 
If the $\bs{R}$ pdf is 3D Gaussian as in the relative diffusion model by 
Batchelor (1952), this factor is only 1.07. Numerical simulations \cite[e.g.,][]{bof02} show 
that the separation pdf of tracer pairs is highly non-Gaussian with 
a very broad tail. The pdf is found to be well fit by the solution 
of the pdf equation proposed by \cite{ric26}. With this broader pdf, 
we find the factor is larger, $\simeq$ 1.20. 
Because $S_{ij}$ and $T_R$ are independent of $R$ for $R \gsim L$, 
our approximation is also expected to be exact for $R$ in 
that range. In conclusion, replacing the $\bs {R}$ pdf by a delta 
function at the rms distance is quite well justified. 
It may overestimate $S_{{\rm T} ij}$ and $\mathcal{D}_{ij}$ 
by $\sim 20$\%, or the relative speed by $\sim 10$\%, if the primary 
distance is in the inertial range (which is the case for $\tau_p$ in the 
inertial range).           

\section{Results}
To calculate the radial relative velocity, we need the 
longitudinal particle structure 
function, $S_{{\rm p} ll}$, which in turn requires 
$\mathcal{A}_{ll}$ and $\mathcal{D}_{ll}$.  
From eq. (2.19), it is clear that $\mathcal{A}_{ll} = \mathcal{A}_{nn}$ and 
\begin{equation}
\mathcal{A}_{ll} =\frac{(\Omega_2 -\Omega_1)^2 \left(\Omega_1 \Omega_2 + (\Omega_1 + \Omega_2) {\displaystyle \frac{z^2}{2} } \right)  }
{(\Omega_1 + \Omega_2) \left(\Omega_1 + \Omega_1^2 + {\displaystyle \frac{z^2}{2}} \right) \left(\Omega_2 + \Omega_2^2 + {\displaystyle \frac{z^2}{2}}\right) } u'^2. 
\end{equation}

We obtain $\mathcal{D}_{ll}$ using the relation $\mathcal{D}_{ll} = \mathcal{D}_{ij} r_i r_j/r^2 $. 
From eq. (2.38) for $\mathcal{D}_{ij}$, we have      
\begin{equation}
\begin{array}{lll}
{\displaystyle {\mathcal D}_{ll}(r) = \int_{-\infty}^0 \frac {d\tau}{\tau_{p1}} \int_{-\infty}^0 \frac {d\tau'}{\tau_{p2}} 
\left[ \left(\frac{1}{3} + \frac{2r^2}{3R^2}\right)S_{ll}(R)  +  \left( \frac{2}{3} -\frac{2r^2}{3R^2}\right) S_{nn}(R)\right]  }\\ \hspace{3.5cm}
{\displaystyle  \times \Phi \big(\tau -\tau'; \tau_{TR}(R), T_R(R) \big) \exp \left( \frac {\tau}{\tau_{p1}} \right)  \exp \left( \frac {\tau'}{\tau_{p2}} \right)}. 
\end{array}
\end{equation} 

We will need the 3D relative velocity under
the assumption of ballistic particle separation. 
In that case, we take the 3D relative velocity 
variance, $\langle w^2 \rangle$, to be $S_{{\rm p} ii} =\mathcal{A}_{ii} + \mathcal{D}_{ii}$ (see \S2.2). 
Using contractions of eqs. (2.19) and (2.38), we have,   
\begin{equation}
\begin{array}{lll}
{\displaystyle \langle w^2 \rangle 
= 3 \mathcal{A}_{ll} + \int_{-\infty}^0 \frac {d\tau}{\tau_{p1}} \int_{-\infty}^0 \frac {d\tau'}{\tau_{p2}} \big( S_{ll}(R) + 2S_{nn}(R) \big ) } 
{\displaystyle \Phi (\tau -\tau'; \tau_{TR}, T_R ) \exp \left( \frac {\tau}{\tau_{p1}} \right)  \exp \left( \frac {\tau'}{\tau_{p2}} \right)}
\end{array}
\end{equation} 
where $\mathcal{A}_{ll}$ is given by eq. (3.1). 

  
\subsection{The Monodisperse Case}
As discussed earlier, for identical particles with 
$\tau_{p1}=\tau_{p2} =\tau_p$, $\mathcal{A}_{ij}$ vanishes 
and only $\mathcal{D}_{ij}$ contributes to the particle 
velocity structure tensor, i.e., $S_{{\rm p} ij}= \mathcal{D}_{ij}$. 
Thus the longitudinal particle velocity structure function 
is given by, 
\begin{equation}
\begin{array}{lll}
{\displaystyle S_{{\rm p}ll}(r) = \int_{-\infty}^0 \frac {d\tau}{\tau_{p}} \int_{-\infty}^0 \frac {d\tau'}{\tau_{p}} 
\left[ \left(\frac{1}{3} + \frac{2r^2}{3R^2}\right)S_{ll}(R)  +  \left( \frac{2}{3} -\frac{2r^2}{3R^2}\right) S_{nn}(R)\right]  }\\ \hspace{3.5cm}
{\displaystyle  \times \Phi \big(\tau -\tau'; \tau_{TR}(R), T_R(R) \big) \exp \left( \frac {\tau}{\tau_{p}} \right)  \exp \left( \frac {\tau'}{\tau_{p}} \right)}. 
\end{array}
\end{equation} 
To solve this equation, one needs the particle dispersion 
laws for $\rho$ to calculate the distance $R$ by eq. (2.24). 
We start with the two simplified cases in order to study the 
effect of each phase in the two-phase separation found 
by \cite{bec09}.  

\subsubsection{Ballistic Separation Behavior} 
We first consider the effect of the ballistic separation 
phase by using eq. (2.35) for the particle distance 
at all times.  
Eq. (3.3) gives the variance of the 3D relative 
velocity, $\langle w^2 \rangle$, needed in eq. (2.35). 
For identical particles, eq. (3.3) becomes,    
\begin{equation}
\begin{array}{lll}
{\displaystyle \langle w^2 \rangle 
= \int_{-\infty}^0 \frac {d\tau}{\tau_{p}} \int_{-\infty}^0 \frac {d\tau'}{\tau_{p}} \big( S_{ll}(R) + 2S_{nn}(R) \big ) } 
{\displaystyle \Phi (\tau -\tau'; \tau_{TR}, T_R ) \exp \left( \frac {\tau}{\tau_{p}} \right)  \exp \left( \frac {\tau'}{\tau_{p}} \right)}.
\end{array}
\end{equation} 
Because the distance $R$ in the integral on the r.h.s. depends on 
$\langle w^2 \rangle$, eq. (3.5) is implicit for $\langle w^2 \rangle$.  
We numerically solve eq. (3.5) by an iterative method. 
After obtaining $\langle w^2 \rangle$, we use it to calculate 
$S_{{\rm p}ll}$ from eq. (3.4).       

The results for the radial relative velocity, 
$\langle |w_r| \rangle$, as a function of the Stokes number, $St$, 
are shown in Fig. 1. We obtained $\langle |w_r| \rangle$ from 
the conversion $\langle |w_r| \rangle = \sqrt{2 S_{{\rm p} ll}/\pi}$ 
assuming a Gaussian distribution for $w_r$. 
We simply follow this convention here 
although we realize that the relative velocity distribution is 
broader than Gaussian for small particles in high 
Reynolds-number flows (e.g., \cite{sun97, wan00}, 
see also \cite{bec09b} for the intermittency in 
inertial particle structures). We will give the relative speed across fixed distances
as a function of Stokes numbers (for convenience in the 
comparison with simulations in \S 3.1.3), although in 
coagulation models one needs to use the relative speed 
across the particle size as a function of the size. 
The latter can be easily calculated from our model 
with the Stokes number as a function of the particle 
size in a specific application.  

In the left panel of Fig. 1, we plot $\langle |w_r| \rangle$ 
for particles at a distance $r=\eta$ as a function of the 
Stokes number and the Taylor Reynolds number. The solution 
reproduces the S-T limit (with the relative speed independent 
of $St$) and the $St^{-1/2}$ scaling in the limit of large friction 
time. As in the model by Zaichik {\it et al}., the relative velocity 
is found to scale as $St^{1/2}$ for intermediate Stokes numbers. 
This scaling corresponds to the inertial-range scaling of the 
turbulent flow and the $St^{1/2}$ scaling range will be referred 
to as the inertial range. 

\begin{figure}
\centering\includegraphics[width=\textwidth]{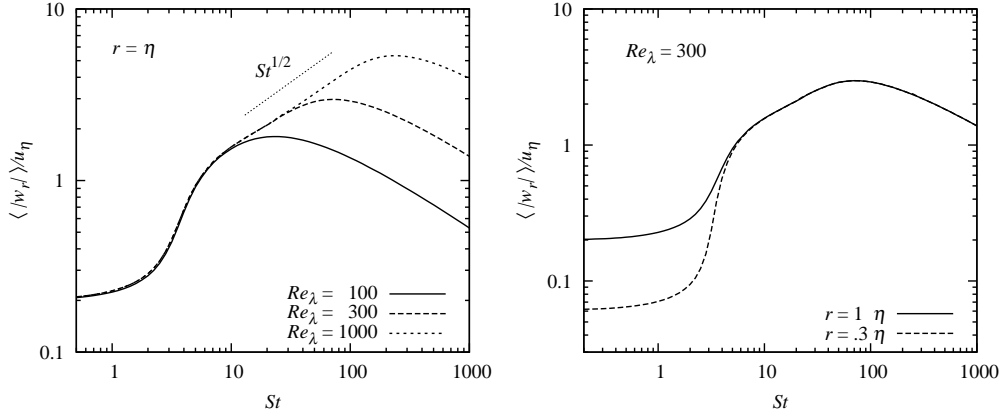}
\caption{The radial relative velocity, $\langle |w_r| \rangle$, 
as function of the Stokes number, $St$, for ballistic particle 
separation. The left panel shows results for different $Re_{\lambda}$. 
The distance $r$ is fixed at $\eta$. A $St^{1/2}$ scaling is found 
between the S-T limit and the large $St$ limit. 
The Stokes number at the transition from $St^{1/2}$ to $St^{-1/2}$ 
scalings increases linearly with $Re_\lambda$. 
The right Panel shows the dependence on 
$r$ with $Re_{\lambda}$ set to $300$. 
} 
\label{f1}
\end{figure}

The $St^{1/2}$ scaling can be understood as follows. 
As discussed earlier, the main contribution to the integrals in 
eqs. (3.4) and (3.5) is from $\tau'$, $\tau \sim -\tau_p$, 
and the particle distance, $R$, at $\tau$, $\tau' \sim -\tau_p$, 
called the primary distance in \S 2.3.3, is important in 
determining the relative velocity. We will denote the primary 
distance as $R_p$. The primary distance as a function of 
$\tau_p$ is evaluated by $R_p = R(-\tau_p, -\tau_p)$ 
using eq. (2.22). For intermediate Stokes numbers,  
$R_p$ is much larger than $r$,  we  thus have 
$S_{{\rm p}ll} = \langle w^2 \rangle/3$ from eqs. (3.4) and (3.5). 
Thus the scaling behavior of $\langle w_r^2 \rangle$ with $\tau_p$ 
is the same as that of $\langle w^2 \rangle$. 
The latter can be obtained by analyzing eq. (3.5). 

The $\Phi$ term in eq. (3.5) represents the persistency 
of a structure of size $R$. It is approximately given by 
$\exp(-|\tau-\tau'|/T_R(R))$. The effect of this factor 
depends on how $T_R(R_p)$ at the primary distance, 
$R_p$, compares to $\tau_p$. If $T_R(R_p)$ is larger 
than $\tau_p$, $\Phi$ would be essentially unity for 
$\tau$ and $\tau'$ values that significantly contribute 
to the integral in eq. (3.5). 
One the other hand, if $T_R$ at $R_p$ is smaller than 
$\tau_p$, a factor of $T_R (R_p)/\tau_p$ needs to be 
accounted 
because the $\Phi$ factor suggests that, for a given 
$\tau$, only $\tau'$ that satisfies $|\tau'-\tau| \lsim T_R$ 
(instead of the range $ -\tau_p \lsim \tau' \leq 0$) 
contributes significantly to the integral. 
We find the latter is the case from our numerical solution.  
Thus, considering the main contribution to the integral is from 
the integrand at $R \sim R_p$, we expect 
$\langle w^2 \rangle \propto [S_{ll}(R_p) + 2 S_{nn} (R_p)] 
T_R (R_p)/\tau_p$. 
For ballistic separation, the primary 
distance $R_p \simeq \langle w^2 \rangle^{1/2} \tau_p$.  
And for $R$ in the inertial range, we have $S_{ll}(R)$, 
$S_{nn}(R) \propto R^{2/3}$ and $T_R(R) \propto R^{2/3}$ 
(eqs. (2.28), (2.32)). These scalings 
give 
$\langle w^2 \rangle \propto \langle w^2 \rangle^{2/3} \tau_p^{1/3}$,  
which results in $\langle w^2 \rangle \propto \tau_p$ and 
hence the $St^{1/2}$ scaling for the relative velocity. 


As the Stokes number increases, the relative speed reaches a peak, 
and transitions to the $St^{-1/2}$ scaling. This change  
occurs when $R_p \simeq L$ and corresponds to the switch 
of the scaling behaviors of $S_{ll}$, $S_{nn}$, and $T_R$ from 
the inertial range to the outer scales. As pointed out earlier, if the 
primary distance, $R_p$, is much larger than $L$, 
the particle velocities are uncorrelated. A $St^{-1/2}$ scaling 
is expected from the particle velocity variances, eq. (2.21).
This scaling can also be obtained using an analysis 
similar to that for the $St^{1/2}$ scaling given above. 
If $R_p \gg L$, the structure functions and the timescale 
$T_R$ that give the primary contribution to the integrals in eqs. 
(3.4) and (3.5) are constant, i.e., $S_{ll}=S_{nn}= 2u'^2$ and $T_R=T_L$ 
(eqs. (2.29) and (2.33)). Therefore, eq. (3.5) gives 
$\langle w^2 \rangle \propto u'^2 T_L/ \tau_p$, 
and hence the $St^{-1/2}$ scaling. Again the factor $T_L /\tau_p$ comes from the $\Phi$ term. 

The Stokes number, $St_m$, or the friction time, $\tau_{pm}$, 
at which the relative speed reaches the maximum can be 
approximately obtained by setting $R_p \sim L$. 
Using $R_p \simeq \langle w^2 \rangle^{1/2} \tau_p$ 
for the ballistic separation and $\langle w^2 
\rangle^{1/2} \simeq 1.0 St^{1/2} u_\eta$ in the inertial range 
from our numerical solution, we find that $St_m \simeq Re_\lambda/\sqrt{15}$, or $\tau_{pm} \simeq T_e$, 
where $T_e$ is defined as $u'^2/\bar{\epsilon}$. 
The order-of-magnitude estimate for $St_m$  turns out to be in good 
quantitative agreement with the numerical solution in  the left panel 
of Fig. 1, which also confirms  the linear increase of $St_m$ with the 
$Re_\lambda$.  


The right panel of Fig. 1 shows the dependence of the 
relative velocity on $r$, the particle distance at time zero. 
For $r \lsim \eta$ of interest here, the relative velocity 
depends on $r$ only in the S-T limit where it increases 
linearly with $r$ (see the shear term in eq. (1.1)). 
For larger particles, it becomes independent of $r$. 
This is because in that case the contribution is mainly 
from the particle memory of the flow velocity difference 
when the particle distance was much larger than $r$.

We have used the same distance, $R$, for the timescale $T_R$ 
in the $\Phi$ term as that for the structure functions. As discussed 
in \S 2.2, for $T_R$ it might be a better choice to use the particle 
distance, $R_m$, at the earlier one ($\min(\tau, \tau')$) of the two 
times $\tau$ and $\tau'$. 
With this choice, $T_R$ is larger because the separation is larger at 
earlier time. The predicted relative velocity is also larger, and we 
find an increase by $\sim$ 30\% in the inertial range.  


\subsubsection{Tracer-like Separation Behavior}
In this subsection, we examine the effect of
the tracer-like separation phase.  
Our main purpose here is to study how the relative 
velocity scales with the Stokes number from the R-O law.  

Here we will consider an initial distance of 
$r=\eta$, and start the separation 
with $\rho(\tau)^2= r^2 + g \bar{\epsilon} |\tau|^3$.  
We switch to the diffusive regime when the separation exceeds $L$ (\S2.3), 
and the connection is chosen such that $\rho(\tau)^2 = L^2 + 2 D |\tau-\tau_d| $
for $|\tau| > |\tau_d|$, where $\tau_d$ is the time 
when $\rho$ reaches $L$.  
In Fig. 2, we show the results for the relative velocity 
as a function of $St$ and $Re_\lambda$ for particles 
at $r=\eta$. The Richardson constant is set to be $g=1.15$. The 
solution correctly reproduces the limits at small 
and large friction timescales as expected, and 
interestingly, we also find a $St^{1/2}$ scaling for 
the relative velocity in the intermediate range of $St$. 

\begin{figure}
 \centering\includegraphics[width=\textwidth]{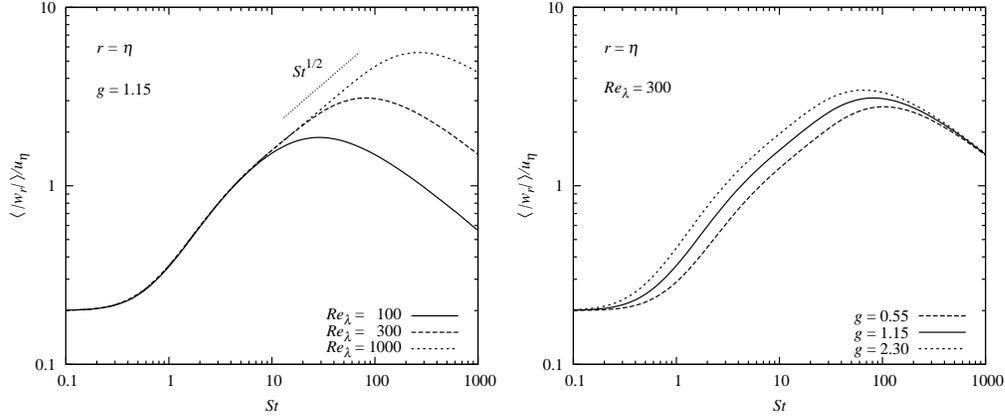}
\caption{The radial relative velocity, $\langle |w_r| \rangle$, 
at $r=\eta$ as a function of the Stokes number, $St$,  
assuming the R-O separation followed by a diffusive phase.  
The left panel shows results for different 
$Re_{\lambda}$. As in the case of ballistic separation, 
a $St^{1/2}$ scaling is found in the inertial range. The Stokes number where the curve 
peaks also increases linearly with $Re_\lambda$. The right panel gives
 the dependence of the relative  velocity on $g$. In the inertial range 
 it increases with $g$ as $g^{1/3}$. 
 }
\label{f2}
\end{figure}

To explain the $St^{1/2}$ scaling,  we analyze again $S_{{\rm p}ll}$ by considering 
the integrand in eq. (3.4) at the primary distance, $R_p$. 
For $R_p \gg r$, eq. (3.4) can be approximately 
written as $S_{{\rm p}ll} \sim [S_{ll}(R_p) + 
2 S_{nn}(R_p)] T_R(R_p)/(3 \tau_p)$. 
The factor $T_R(R(\tau_p))/\tau_p$ (which is 
smaller than 1 for any $g \lsim 10$) is based on 
the same reasoning as in \S3.1.1 for the 
ballistic case. 
With the R-O separation law, the primary distance is given by 
$R_p \simeq (g \bar{\epsilon} \tau_p^3)^{1/2}$. Therefore, 
using the inertial-range 
scalings of $S_{ll}$, $S_{nn}$,  and $T_R$, we have 
$S_{{\rm p}ll} \propto g^{2/3} \tau_p$. 
This explains the $St^{1/2}$ scaling for the radial relative velocity 
in the inertial range, and also predicts that the relative velocity 
increases with the Richardson constant as $g^{1/3}$. 
The $g^{1/3}$ dependence is confirmed by the right panel 
of Fig. 2. 
The increase of the relative velocity with $g$ shows that faster 
particle separation gives larger relative velocity.

As in the ballistic case, 
the friction time, $\tau_{pm}$, at which the relative 
velocity peaks, is again obtained by setting the 
primary distance $R_p$ to $L$. Using the R-O 
separation law for $R_p$, we find that 
$\tau_{pm} \simeq g^{-1/3} T_e$, with $T_e = u'^2/\bar{\epsilon}$. 
In units of the Kolmogorov timescale, we have $St_m \simeq g^{-1/3} 
Re_\lambda/\sqrt{15}$. This is consistent with the results in the left panel of 
Fig. 2. 

We have finished the study for the effects of each separation 
phase in the two-phase separation found by \cite{bec09}.  
We found that, in the inertial range, the relative speed from 
using the R-O law has the same scaling behavior 
as from the ballistic separation. Thus the same $St^{1/2}$ scaling is 
also expected for a combination of ballistic and R-O separation 
behaviors. For $g \sim 1$, the predicted value for the relative speed 
in the inertial range from the R-O separation is quantitatively 
very close to that in the ballistic case, and so is the peak Stokes 
number $St_m$. Therefore for the combined behavior the prediction 
for the  inertial range and for the switch to large Stokes numbers 
would be similar to the two simplified cases (see Fig. 4). 
However, in the transition region from the S-T limit to inertial range, 
the predicted relative speed is quite different for the two separation 
behaviors.   

\subsubsection{Combined Separation Behavior and Comparison with Simulation Results} 
We use the simulation results by \cite{wan00} to test our model. 
The Reynolds numbers in their simulations are quite low with 
$Re_{\lambda}$ in the range from 45 to 75.  The results by \cite{wan00} 
for three different Reynolds numbers are shown as data points in Fig. 3. The relative speed 
is for particles at a separation of $\eta$.  Due to the limited resolution of 
these simulations, no inertial-range scaling is seen. For $Re_{\lambda}$ 
in the same range,  our model with the two 
separation behaviors considered above  does not show the 
$St^{1/2}$ scaling either (see $Re_{\lambda} =100$ curves in Figs. 1 and 2).

The separation behavior of inertial particle pairs at Reynolds numbers  
as those in the simulations by \cite{wan00} is unknown. We tried 
different separation behaviors 
and compared the predictions for the relative speed with their results. 
We find that    
a combination of an early ballistic phase and a later 
tracer-like phase, as found by \cite{bec09}, 
can give a quite good fit to the results by \cite{wan00} (while 
with a pure ballistic separation or a pure tracer-like separation 
no satisfactory fit is found). The lines in Fig. 3 show the predicted 
relative velocities with such a two-phase separation, which 
agree quite well with the simulation data. The exact separation 
behavior used in Fig. (3) is as follows.  The separation starts 
with a ballistic phase which lasts from time zero back to 
$\tau_c$.  At  $\tau_c$, it continuously connects to the 
R-O separation law,  and finally switches to the diffusive 
regime when the separation exceeds $L$. The connection 
between the ballistic phase (eq. (2.35)) and the R-O phase (eq. 2.36) 
is chosen such that the particle distance at $\tau <  \tau_c$  is given by 
$\rho(\tau)^2  = \rho(\tau_c)^2 + g |\tau -\tau_c| \tau^2$. 
This connection between the two phases is quite smooth.  
Furthermore, we set $\tau_c = -1.4 \tau_p$ and  $g= 1$. 
The connection between the R-O regime and the diffusive 
regime is the same as that for the tracer-like separation behavior discussed in \S 3.1.2.  
Due to the low Reynolds numbers here, 
there is only a very short period for the R-O separation between 
the ballistic regime and the diffusive regime in the chosen separation behavior.   
How the ballistic phase exactly connects to the R-O phase 
in the backward dispersion of inertial particles is unknown.  
We also tried other ways to connect the two phases,  and 
found that, by adjusting $\tau_c$ and $g$, some other 
connections can also give satisfactory fits. For example, 
if the two phases are connected by $\rho(\tau)^2  = \rho(\tau_c)^2 + g (\tau -\tau_c)^2|\tau|$ 
for $\tau < \tau_c$, fitting the simulation data gives 
$\tau_c = -1.1 \tau_p$ and $g = 1.5$; and for a connection 
with $\rho(\tau)^2  = \rho(\tau_c)^2 + g |\tau|^3 -g|\tau_c|^3$ 
in the R-O phase, we find a good fit 
with $\tau_c = 1.5 \tau_p$ and $g=0.6$. 

\begin{figure}
\centering\includegraphics[width=0.9\myfigwidth]{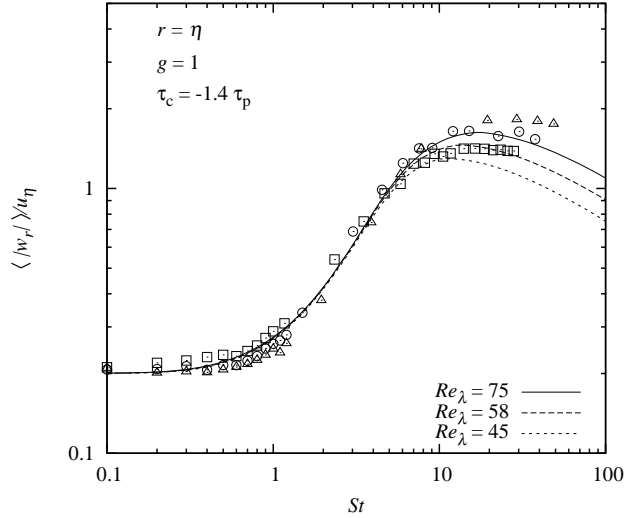}
\caption{Comparison with the simulation results from \cite{wan00} 
at $Re_{\lambda}= 45 $ (squares), 58 (circles) and 75 (triangles). 
Lines are the predicted relative speed from our model with 
a two-phase separation behavior,  a ballistic phase
 followed a tracer-like phase. The ballistic separation is assumed to 
connect with the R-O separation with $g=1$ at $-1.4\tau_{\rm p}$.  
See text for details on this connection.}  
\label{f3}
\end{figure}

Because $\mathcal{D}_{ij}$ is an integral over the history of the particle distance,
different separations as a function of time could lead to the same 
predicted relative speed. Therefore, the exact backward 
dispersion behavior  cannot  be determined by fitting the simulation results 
for the relative speed. In other worlds, a separation law that fits the 
data may not represent the exact dispersion behavior for inertial 
particles in the simulated flows by \cite{wan00}, and a verification would 
need a direct numerical study of the pair separation. However, the fact that 
the dispersion behavior used in Fig. 3 is generally consistent with 
the simulations results by \cite{bec09} (see their Fig.8 where 
the connection between the ballistic phase and the R-O phase 
occurs between 1-2 $\tau_p$ ) suggests that the adopted behavior 
is at least qualitatively correct. We will adopt the separation behavior 
used in Fig. 3 for all the calculations in the rest of the paper.  

We find a significant deviation ( up to 25 \%) 
between the model prediction and the simulation results 
for very large particles ($St \gsim 20)$. It seems that the 
deviation at these large Stokes numbers could not be 
removed by a reasonable change in the separation 
behavior without causing discrepancy at smaller $St$. 
This deviation also occurs in Zaichik {\it et al.}'s model.  
An immediate suspect for this deviation is the assumption 
in our model (and in Zaichik {\it et al.}'s model) that the trajectories 
of all particles are not far away from those of the fluid elements  
This assumption is not well justified for very large particles. 
As discussed at the end of \S 2.2, the temporal correlation of the flow 
velocity (or the velocity difference) along the trajectories of 
large particles may be close to the Eulerian correlation,  
while we used the Lagrangian correlation timescale throughout the 
model.  If the Eulerian correlation 
timescale were used for these particles, and if the Eulerian correlation 
timescale is larger than the Lagrangian timescale 
(see discussions in \S2.2),  the predicted relative speed for large particles 
would be larger, reducing the the difference between the model
and the simulation results. 
Our approximation for the trajectory structure tensor in \S 2.2 
could also contribute to the discrepancy.

\begin{figure}
\centering\includegraphics[width=0.9\myfigwidth]{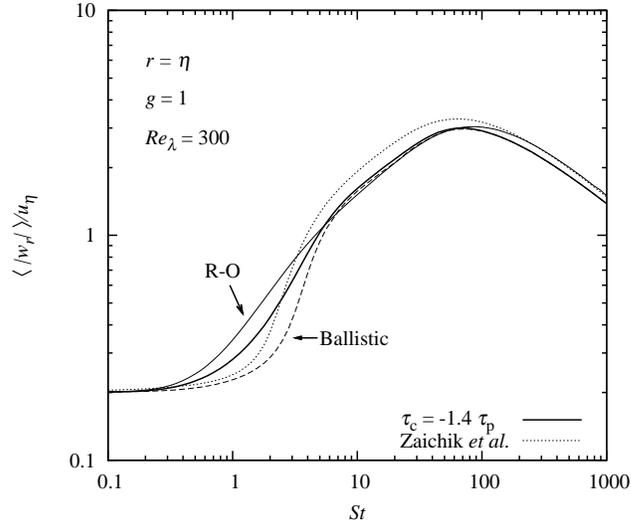}
\caption{The radial relative velocity, $\langle |w_r| \rangle$, 
as a function of the Stokes number, $St$,  with the same 
two-phase separation used in Fig. 3. The dashed thin line is for
ballistic separation only and the solid thin line is for tracer-like 
behavior only (with $g=1$ for the R-O separation). 
Also shown is the prediction by the Zaichik {\it et al.}'s model.}
\label{f4}
\end{figure}

Using the separation behavior in Fig. 3 that well fits the simulation 
results, we carried out calculations for larger Reynolds numbers.  
Fig. 4 shows our prediction for $Re_\lambda =300$.  
For comparison, we included results assuming pure ballistic 
separation (the dashed thin line) and pure 
tracer-like separation (the solid thin line). For the latter we used  $g=1$ 
for the R-O separation.  We see that, in the two-phase case, the relative 
velocity in the transition region from the S-T limit to the inertial range lies 
between the two cases with single separation behavior. In this 
transition region,  the predicted relative speed relies on how 
the ballistic phase and the R-O phase exactly connects.   
In the inertial range, we again have the $St^{1/2}$ scaling, which is 
expected because both the ballistic behavior and the R-O 
separation law give that scaling. 
For $g \sim 1$, the predicted relative speed  turns out to be very similar
for all the three cases in the inertial range. 
The predicted speed in the inertial range is not 
significantly affected by the details of the connection between 
the ballistic and the R-O phases.  The Stokes number 
at the transition from the inertial range to the $St^{-1/2}$ 
range is  $\sim Re_\lambda/\sqrt{15}$ ($\tau_p \sim T_e$), 
similar to both the pure ballistic case and the pure tracer-like case.  



Fig. 4 also shows the relative speed predicted by Zaichik {\it et al}.'s model 
(the thin dotted line). We obtained the results of their model by numerically 
solving the set of differential equations given in \cite{zai06} (i.e., their eqs. (51-53)). 
The transition from the S-T limit to the inertial range in Zaichik et al.'s 
model is quite steep.  The relative speed in the inertial range in their 
model is 20\% larger than the our model.  This is probably because 
a faster particle separation is built into Zaichik {\it et al}.'s model.                   
As pointed our earlier,  in the limit of $\tau_p \to 0$, Zaichik {\it et al}.'s 
formulation implies a Richardson constant significantly larger than 1. 
  



\subsection{The Bidisperse Case} 
We first point out the importance of gravity in the bidisperse 
case. In the monodisperse case, the settling velocity by 
gravity is the same for all particles, and thus neglecting 
gravity in that case may give approximately good estimates 
for the relative speed. The situation is quite different in the bidisperse 
case where gravity may play a major role, especially for large 
particles. The terminal velocity difference between two different particle 
could give substantial or even dominant contribution to their 
relative speed.  
Besides the direct contribution to the relative speed, 
differential settling can also have an indirect 
effect by increasing the particle separation in the past. Since a larger 
separation backward in time would increases the contribution 
from the particle memory of the flow velocity difference,  
this indirect effect also tends to give a larger collision speed.  

We will neglect gravity in our model below.  
Clearly, this limits the application of the model only to 
situations where the relative speed caused by 
differential settling is negligible in comparison 
to the prediction of our model.  However, understanding 
the simpler case with turbulence alone is of theoretical 
importance because it serves as the first step to a physical 
and accurate model for particle collisions in realistic environments 
where both turbulence and gravity are present. The 
effect of gravity may be included in our framework by 
accounting for both the direct effect of differential settling  
and its indirect effect through the backward separation of the 
particles (e.g., in a similar way as  in \cite{aya08}, who, however,
did not include the particle separation by the turbulent flow). 

In the bidisperse case, we include the contribution from the $\mathcal{A}_{ll}$ term,  
eq. (3.1), to the relative speed.  We use the same numerical method to solve $\mathcal{D}_{ll}$ as in the 
monodisperse case. The particle separation behavior is chosen to be similar to the one that well 
fits the simulation results in the monodisperse case.  Note that in the ballistic phase, the 
separation speed  (i.e., the relative velocity variance, $\langle w^2 \rangle$) to be used for 
the calculation of $\mathcal{D}_{ll}$ has a contribution from the $\mathcal{A}_{ii}$ term (eq. 3.3).  
It is not clear how long the ballistic phase lasts in the bidisperse case,  
since the friction timescales of the two particles are different. 
We simply assume that the duration of the ballistic separation is proportional to the 
average of the two friction timescales. Fig. 5 shows the radial relative velocity as a 
function of the Stokes number of particle (2) for a fixed Stokes number, $St_1 =1$, of particle (1).
The data points are simulation results from \cite{zho01} 
for $r= \eta$ at  $Re_\lambda=45$ and $58$. These data are from Fig. 15 of  
\cite{zho01}, but a different normalization is used here. The lines are 
the prediction of our model where we adopted the same connection used in 
Fig. 3 for the monodisperse case and set $g=1$ and $\tau_{\rm c} = - 1.4 \times (\tau_{\rm{p1}} + \tau_{\rm{p2}})/2$.
The agreement of the model prediction with the simulation results is quite good, except that it is a 
little bit broader around the dip. A possible reason is that the separation behavior for different 
particles is different from the one used here,  which is based on the separation behavior of 
identical particle pairs.

\begin{figure}
\centering\includegraphics[width=0.9\myfigwidth]{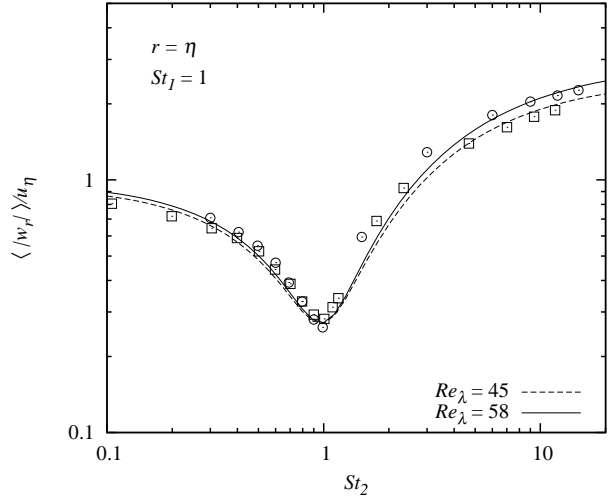}
\caption{Comparison with the simulation results from \cite{zho01} 
at $Re_{\lambda}= 45 $ (squares), 58 (circles). 
Lines are the predicted relative speed from our model with 
a two-phase separation behavior,  a ballistic separation followed by
a tracer-like behavior. The ballistic separation is assumed to 
connect with the R-O separation with $g=1$ at $-1.4\tau_{\rm p}$.  
See text for details about the connection.}  
\label{f5}
\end{figure}


We give more details for the relative speed between different particles in Fig. 6,   
where the Taylor Reynolds number is set to $300$. 
The thick lines are for $r= \eta$.  A dip around $St_2 \sim St_1$ is 
found in every curve with $St_1 \lsim 100$. The existence of  
the dips is related to the fact that the contribution from $\mathcal{A}_{ij}$ for 
particles of similar sizes is small. Physically, it means that the velocities of 
similar particles tend to have stronger correlation than 
particles with very different friction timescales. 
Around each dip, the contribution to the 
relative velocity is mainly from $\mathcal{D}_{ll}$, while far from the dip 
it is dominated by $\mathcal{A}_{ll}$. 
A comparison of the two terms shows that they give 
similar contributions when the Stokes number ratio 
is about 3-4. If the friction timescales of the two 
particles differ by a factor much larger than 4, using the 
generalized acceleration term alone may give a satisfactory 
result. The $\mathcal{D}_{ll}$ and $\mathcal{A}_{ll}$ 
terms in our model are closely related to the 
two terms in the equation for the velocity 
difference given in \cite{bec05} (their equation (13)). 
Their discussion on the relative importance of those two 
terms provides physical insights to understand the 
dips in Fig. 6. The relative speeds at the dip centers 
correspond to equal Stokes numbers. Connecting these 
centers would give a curve identical to that for the monodisperse 
case with the same separation behavior and parameters.   

\begin{figure}
 \centering\includegraphics[width=0.9\myfigwidth]{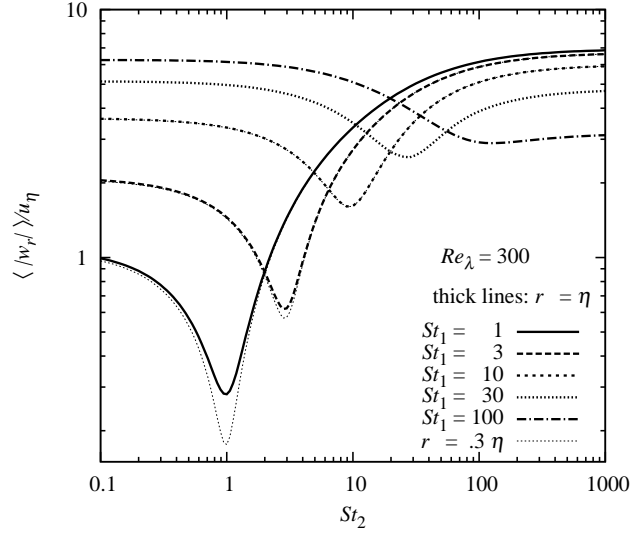}
\caption{The radial relative velocity, $\langle |w_r| \rangle$ 
as a function of $St_2$ with fixed $St_1$. The thick lines  
give the relative speed at $r= \eta$. Each curve has a 
dip at $St_2 = St_1$. The three thin dotted lines show the 
results for $r = 0.3 \eta$ with $St=1, 3,$ and 10.   
}
\label{f6}
\end{figure}

On the far left of the dip, the relative velocity approaches a 
constant. The constant corresponds to $\mathcal{A}_{ll}$ 
in the limit $\Omega_2 \to 0$, which, from eq. (3.1), is given by
$\frac{\Omega_1^2}{\Omega_1 + \Omega_1^2 +z^2/2}u'^2$. 
As $St_1$ (or $\Omega_1$) increases, the dip moves to 
the right, and the relative velocity on the far left increases. 
It reaches and stays at the maximum (corresponding to 
$\mathcal{A}_{ll} =u'^2$) after $\Omega_1$ 
becomes much larger than 1. The opposite occurs 
on the far right of the dip, i.e., in the limit $St_2 \gg St_1$. 
In this limit, $\mathcal{A}_{ll}$ approaches 
$\frac{\Omega_1 +z^2/2}{\Omega_1+\Omega_1^2 +z^2/2} u'^2$, 
which has a maximum of $u'^2$ at small $St_1$. With increasing 
$St_1$ or $\Omega_1$, the relative velocity decreases on the 
far right of the dip, while it increases on the other side.     

The dip disappears for very large $St_1$ 
(larger than $\sim 100$ for the case shown in the Fig. 6). 
This can be explained as follows. Physically, the dips are 
due to strong velocity correlation between particles of 
similar sizes. Thus no dip would exist if the friction 
time of particle (1) is such that its velocity is not significantly 
correlated with any particle of similar size. This is the case 
for a particle with $\tau_p$ larger than $\tau_{pm}$, the friction 
timescale where the relative speed peaks in the monodisperse case (see \S 3.1).  
The velocity of such a particle is not correlated even with 
an identical particle. Therefore, no dip is expected around 
$St_2 \simeq St_1$ if $\tau_{p1} \gsim \tau_{pm} \simeq T_e$.     
Clearly, this argument suggests that the critical value 
of $St_1$ where the dip starts to disappear is the 
same as the Stokes number where the relative 
speed peaks in the monodisperse case. 



The dependence of the relative velocity on the 
distance $r$ for the bidisperse case is also illustrated 
in Fig. 6,  where the dotted thin lines show results for $r= 0.3 \eta$. 
The $r$ dependence only comes from $\mathcal{D}_{ll}$, 
because $\mathcal{A}_{ll}$ does not depend on $r$. Thus the 
dependence may exist only around dips where $\mathcal{D}_{ll}$ 
gives a significant contribution. As in the monodisperse case, 
$\mathcal{D}_{ll}$ is independent of $r$ (again for $r \lsim \eta$) 
if the Stokes numbers of both particles are much larger than $1$. 
This suggests that,  in the bidisperse case, the relative velocity 
is a function of $r$ only when $St_2 \simeq St_1 \lsim 1$.  
In Fig. 6 we see  that for $St_1 = 1$ the depth of the dip 
increases with decreasing $r$, corresponding to the decrease 
of the relative velocity with $r$ in eq. (1.1) for small identical 
particles.  The $r$-dependence is already weak for $St_1 =3$,  
and the relative velocity becomes completely independent of $r$ for $St_1 =10$.

In summary, we find that in the bidisperse case 
the dominant contribution to the relative velocity is 
from $\mathcal{D}_{ll}$ if the ratio of the two Stokes numbers 
is not larger than 3-4, or from $\mathcal{A}_{ll}$ if that ratio 
is larger. For a fixed Stokes number of one particle, $St_1$,  
the relative velocity as a function of the Stokes number of the 
other particle, $St_2$, shows a dip at $St_2 \sim St_1$. The 
dip corresponds to a stronger velocity correlation between 
particles of similar sizes than between very different particles.
The existence of a dip at equal Stokes numbers      
has consequences for the collision kernel \cite[e.g.,][]{bec05, zai06}. 

\section{Comparison with Other Models}
Our model has already been compared with some previous models 
earlier in the paper. Here we discuss more models and present 
the comparisons more systematically.     

\subsection{Volk et al. (1980)}

In the astrophysical literature, the relative velocities of dust grains, e.g., 
in the context of the coagulation growth of dust grains in protoplanetary disks, 
has almost always been estimated based on the model by \cite{vol80}, 
and a later version of that model by \cite{mar91}. Here we discuss this 
model for the particle relative velocity and compare it with the physical 
picture presented in this paper.

Volk {\it et al} (1980) derived  both the 1-particle 
rms velocity and the relative speed between two 
particles. Their model started by considering  the 
effects of turbulent eddies of different sizes on a particle 
of a given friction timescale. They speculated 
that the effect of an eddy on the particle depends 
on the eddy size. The particle would basically move 
along with large eddies if the eddy turnover time 
is much larger than the particle friction timescale. 
On the other hand, the effect of eddies with turnover 
time much smaller than the friction time is argued to be like 
a ``random kick" because the eddy would ``die" 
within a friction timescale. Apparently ``random kick" here 
only means that the particle does not ``receive" the driving 
by these small eddies to a full extent. There is also another 
way that an eddy 
may behave like a random kick. Due to their inertia, 
particles have a different velocity from the flow.  
If the relative velocity between a particle and a 
turbulent eddy of a given size is such that  the particle 
crosses and leaves the eddy within a friction timescale, 
then the particle would not receive a ``full" kick from that 
eddy, and the effect of the eddy is a ``random kick". 
Based on these considerations, Volk {\it et al.} 
defined a critical eddy size in Fourier space,  $k^*$, 
such that the effect of eddies below this size is like random kicks.

 
In the model by Volk {\it et al.} (1980) for the rms velocity of a single particle, 
the wavenumber $k^*$ appears to be important because eddies of 
size smaller than the scale $l^*$, corresponding to $k^*$, are expected 
to be less efficient at ``driving'' particle motions than larger ones 
because the particle does not have chance to "fully" receive the energy 
from these eddies. 


Volk {\it et al.}'s derivation for the relative speed between two particles
did not consider the separation of two particles, which is essential 
in our physical picture. In their calculation for the relative speed, 
they continue to take $k^*$ as a crucial scale. 
They essentially assumed the velocities of two particles 
induced by eddies with wavenumber larger than $k^*$ 
are not correlated, and the contribution to the particle velocity correlation 
is only from larger eddies.  This assumption is not justified.  

We argue that it is unlikely that $k^*$ is crucial  for the velocity 
correlation of two particles (although $k^*$ may be an important 
scale for the 1-particle rms velocity, as discussed above). 
Whether the particle relative motions induced by eddies at a given 
scale are correlated or not is probably determined by 
how the particle separation compares to the eddy size. If the distance 
of two particles is smaller than the size of an eddy they encounter, the 
particle motions induced by this eddy would be correlated 
even if the eddy size is smaller than $l^*$ (the contribution 
to the two particles from a same eddy should be correlated).  
In this case, Volk {\it et al}.'s assumption would 
underestimate the correlation and  overestimate the relative velocity. 
On the other hand, contributions to the velocities of the 
two particles from eddies of size smaller than the particle
distance would be uncorrelated, because each particle 
receives a contribution from a different eddy of that size and 
motions in different eddies are likely to be independent. Therefore, contrary to 
the assumption by Volk {\it et al}., particle motions 
induced by eddies of size larger than $l^*$ are not always 
correlated. They are independent if the size of these eddies 
is smaller than the particle separation. The argument above suggests that it is the particle 
separation (instead of $l^*$ or $k^*$) that determines the particle velocity correlation, and 
thus an explicit examination of the particle separation is required. 

The model by Volk {\it et al.} for the relative speed for two particles may  
be interpreted as one  that implicitly 
assumes that the typical particle separation, $R$, is around the scale $k^*$. 
However, this assumption cannot  be physically justified since the definition 
of $k^*$ has nothing to do with the distance between two particles. 
Even if the value of $k^*$ turned out to be close the to typical distance between particles, 
it should probably be taken as a coincidence. Volk {\it et al.}'s model may be improved by 
incorporating the particle distance as a function of time within their 
formulation.

\subsection{\cite{wil83}} 
\cite{wil83} started from the derivation of the 
relative velocities for particles in two limits: $\tau_p \ll T_L$ 
and $\tau_p \gg T_L$, and then obtained a ``universal" formula 
by interpolation. In the limit of $\tau_p \ll T_L$, 
they assumed that the particle separation back in time can 
be neglected (\S 2.2), and found the relative velocity is given 
by eq. (2.20). As argued in \S 2.2, this result is not valid 
for similar particles especially for $St \gg 1$. 
In the other limit of $\tau_p \gg T_L$, Williams and Crane (1983)
considered a linear particle separation, and chose 
the separation rate assuming the particle velocities are not 
correlated in this limit. 

\cite{wil83} found a ``universal" formula
that reproduces the results for two limits they considered.    
This formula is obtained from a mathematical interpolation, 
and thus does not incorporate the physics of the relative 
velocities between particles of intermediate inertia. We find that the 
formula gives a $St^{3/2}$ scaling for identical particles 
with $\tau_p \lsim T_L $, which is probably incorrect.  
    
\subsection{\cite{yuu84}}
\cite{yuu84} derived a formula for the the relative velocity 
that consists of an acceleration term and a shear term. 
Neglecting the effect of the added mass term (which is 
negligible in gaseous flows) included in Yuu's calculations, the 
acceleration term is also given by eq. (2.20), i.e., exactly the 
same as Williams and Crane's result for the small particle limit. 
Yuu's shear term is much smaller than our generalized shear term,
$\mathcal{D}_{ij}$, for intermediate to large particles, because 
his calculation did not consider the particle separation back 
in time, and thus the shear term does not account for particles' 
memory of the larger flow velocity difference they ``saw" 
at earlier times. As pointed out in \S.2.2, this leads to 
a significant underestimate of the contribution from 
$\mathcal{D}_{ij}$. As a consequence, Yuu's model is not 
valid for similar particles with $ \tau_p \gsim \tau_\eta$.   

\subsection{\cite{kru97}} 
\cite{kru97} gave a generalization to 
the models of \cite{wil83} and 
\cite{yuu84}. They first noticed that Williams and 
Crane's result for particles with $\tau_p \ll T_L$ 
and Yuu's acceleration term (eq. (2.20)) 
do not reproduce that in eq. (1) for the S-T limit. 
Replacing the temporal energy spectrum in \cite{wil83} by 
one that incorporates the flow acceleration 
and corresponds to a temporal correlation function 
similar to our eq. (2.16), they were able to derive a 
formula for the $\tau_p \ll T_L$ case that correctly 
reduces to the acceleration term in the S-T limit.  
The formula is the same as our result for $\mathcal{A}_{ij}$ 
in the limit $z \ll 1$. \cite{kru97} also 
generalized the Williams and Crane model to include the 
added mass effect, which was considered in \cite{yuu84}. 
The effect is negligible in a gaseous flow, but may be 
important in liquid flows. Following Williams and Crane, a 
``universal" solution was obtained by interpolating 
the generalized results for the two limits. Therefore 
the Kruis and Kusters model shares the same weakness 
as \cite{wil83}: the physical importance 
of the particle separation for the relative velocity of 
similar particles with $\tau_p \lsim T_L$ is not 
included in the model, and the interpolated results 
for that case are probably incorrect.  

\subsection{The analytical model of Zaichik et al. (2003, 2006)} 
In addition to their differential model discussed 
earlier in details, Zaichik and collaborators also 
presented an analytical model. Assuming Gaussian 
statistics for both the flow and the particle 
velocities, the analytical model calculates 
the joint pdf of the velocities of two particles, 
$P(\bs{v}^{(1)},\bs{v}^{(2)})$, from the two-point 
joint pdf of particle and flow velocities, 
$P(\bs{v}^{(1)},\bs{v}^{(2)},\bs{u}^{(1)},\bs{u}^{(2)} )$. 
\cite{zai03, zai06} approximated the latter by  
$P(\bs{v}^{(1)}|\bs{u}^{(1)}) P(\bs{v}^{(2)}|\bs{u}^{(2)} ) 
P(\bs{u}^{(1)},\bs{u}^{(2)})$ 
under the assumption that 
$P(\bs{v}^{(1)}|\bs{v}^{(2)},\bs{u}^{(1)},\bs{u}^{(2)} ) 
= P(\bs{v}^{(1)}|\bs{u}^{(1)})$ and 
$P(\bs{v}^{(2)}|\bs{u}^{(1)},\bs{u}^{(2)} ) = 
P(\bs{v}^{(2)}|\bs{u}^{(2)} )$. 
This assumption is valid in the limit $\tau_p \to 0$. The particle 
velocity is well approximated by the flow velocity at the same 
point in this limit so that $P(\bs{v}^{(1)}|\bs{v}^{(2)},\bs{u}^{(1)},\bs{u}^{(2)} ) 
\simeq  \delta (\bs{v}^{(1)}-\bs{v}^{(2)})$.  
Therefore it can be approximated by $P(\bs{v}^{(1)}|\bs{v}^{(2)})$, which is  
also $\delta (\bs{v}^{(1)}-\bs{v}^{(2)})$ in the limit.  A similar argument applies to 
$P(\bs{v}^{(2)}|\bs{u}^{(1)},\bs{u}^{(2)} ) = P(\bs{v}^{(2)}|\bs{u}^{(2)} )$. 
In the other limit with $\tau_p \gg T_L$, we find the assumption for the 
conditional pdfs is also roughly valid.  In this limit, due to  the long memory,  
the particle velocity, $\bs{v}^{(1)}$ is not strongly correlated with the local flow 
velocity  $\bs{u}^{(1)}$, nor with $\bs{u}^{(1)}$ or $\bs{u}^{(2)}$. 
Therefore $P(\bs{v}^{(1)}|\bs{v}^{(2)},\bs{u}^{(1)},\bs{u}^{(2)} ) 
= P(\bs{v}^{(1)}|\bs{u}^{(1)})$ could be a good approximation in the large particle limit.  
(The predicted relative velocity by this model could reproduce the two limits given in \S1). 
However, for $\tau_p$ in the inertial range, all the four velocities are 
partially correlated. The assumption that neglects the ``direct" correlation of the 
particle velocity at point (1) with the flow and particle velocities at point (2) would, 
to some degree, 
underestimate the particle velocity correlation. 


From the 2-point joint pdf of particle and 
flow velocities, \cite{zai03} derived the joint pdf 
$P(\bs{v}^{(1)},\bs{v}^{(2)})$ and an analytical formula 
for the relative velocity. In the derivation, the Lagrangian 
correlation function is needed to calculate the velocity 
variance of each particle and the flow-particle 
correlation at each point, which fix 
$P(\bs{v}^{(1)}|\bs{u}^{(1)})$ and 
$P(\bs{v}^{(2)}|\bs{u}^{(2)})$ under the 
assumption of Gaussian statistics. With our 
eq. (2.16) for the Lagrangian correlation function, 
the model gives a linear scaling with $St$ in 
the inertial range for the monodisperse case. 
This is in contrast to the $St^{1/2}$ scaling in 
both their differential model and our model. Because 
the analytical model underestimates the velocity 
correlation, the predicted relative velocity for 
identical particles in the inertial range is much 
larger than in the latter two models, as well as 
than the results from simulations with low 
Reynolds numbers \cite[][]{zai03}. Furthermore,  
the analytical model does not give 
a dip at $St_2 \simeq St_1$ in the bidisperse case. 
This is again because the velocity correlation between 
similar particles is not accurately evaluated. 
The model does not account for the fact that 
similar size particles tend to have stronger correlation 
than different size ones.   

\subsection{\cite{der06}}  
The model of \cite{der06} starts with a similar approach 
as in the differential model of Zaichik {\it et al}. An 
equation for the joint pdf of both positions and velocities 
of the two particles is derived assuming Gaussian 
statistics for the flow velocity field.  
The equation is equivalent to that for 
the joint pdf of the particle separation 
and the relative velocity in Zaichik {\it et al}.'s model. 
Apparently, Derivich (2006) considered the solution 
of the joint pdf equation only in the zero separation 
limit \footnotemark\footnotetext{
Zaichik {\it et al}. 
considered the whole range of the particle separations 
when solving their joint pdf equation 
and thus were able to  examine the spatial clustering of particles that 
Derevich's model does not address.}.  Neglecting the spatial derivative 
terms in the joint pdf equation (i.e., assuming the spatial derivatives can 
be neglected in the zero separation limit), Derivich obtained a solution 
for the joint pdf (Gaussian) of the velocities of the two particles. 
The correlation of the particle velocities in the solution depends on 
the particle separation as a function of time (with initial separation 
set to be zero). In his calculations, the particle separation is taken to 
be a Gaussian variable. The time 
dependence of the separation variance is neglected, and apparently 
for the monodisperse 
case the variance is set to be a constant,  corresponding to the 
particle distance at $\tau$, $\tau' =-\tau_p$ in the ballistic separation 
behavior assumed in 
our model. We argue that this treatment with a constant particle separation 
variance is physically inadequate.   
     

\subsection{Ayala et al. 2008}
Our formulation is very similar to that in \cite{aya08}. The model 
by \cite{aya08} included the 
particle separation due to gravity for sedimenting droplets 
in turbulent flows, but neglected the particle separation 
by turbulent dispersion. In the absence of gravity, 
particles do not separate in their model, and  the model 
enormously underestimates the relative velocity in 
the monodisperse case (where the separation plays a crucial role), 
as can be seen from their Fig. 11b. As explicitly pointed out 
by \cite{aya08}, their model was not designed for 
particles of similar sizes. 
In the presence of gravity, it is expected the  
accuracy of the model decreases with increasing turbulence 
intensity. 
This can be seen from their fig. 13 for the monodisperse case, 
where  the predicted relative velocity agrees with the simulation 
results for the lower of the two turbulent intensities 
shown ($\bar{\epsilon} = 100$ cm$^2$/s$^3$), while it is 
significantly smaller than the simulation results 
for the case with the higher intensity ($\bar{\epsilon} = 400$ cm$^2$/s$^3$).  
Clearly, with larger turbulent intensity, the turbulent 
dispersion is faster and neglecting it would result in 
less reliable predictions. 
We also note that Ayala {\it et al}. adopted a bi-exponential form
(similar to the form of our eq. (2.16) for the Lagarangian 
temporal correlation function) for the spatial correlation 
function of the flow velocity. The form corresponds to a linear 
velocity difference scaling and hence a $-2$ energy spectrum in the inertial 
range (for comparsion, see eq. (2.29) for the spatial correlation 
function adopted in our model), and is thus not consistent with 
the Kolmogorov spectrum observed in turbulent flows of high 
Reynolds numbers.       

\section{Conclusions}  
We have examined the relative velocity of inertial particles 
suspended in turbulent flows. A general formulation 
is established based on the calculation of the 
particle velocity structure function. Our general result 
for the particle structure function, eq. (2.12), has two terms, 
a generalized acceleration term, $\mathcal{A}_{ij}$ (eq. (2.13)), 
and a generalized shear term, $\mathcal{D}_{ij}$ (eq. (2.14)). 
The generalized shear term, $\mathcal{D}_{ij}$, 
corresponds to the contribution to the relative speed 
from particles' memory of  the flow velocity difference 
in the past.  We find that the backward-in-time dispersion 
of inertial particle pairs is needed to evaluate this term. 
The two terms reduce to the acceleration term and 
the shear term, respectively, in the S-T limit. 
Our formulation can thus be viewed as a generalization 
of Saffman and Turner's result for the limit of 
small particles to particles of any size. 




We have shown that our model with a separation behavior 
similar to that found by recent simulations for the forward 
(in time) dispersion of inertial pairs \cite[][]{bec09}, i.e., 
a ballistic separation followed by a tracer-like behavior, 
gives quite good fits to the relative speed measured 
from simulations by \cite{wan00} (for the monodisperse case) 
and by \cite{zho01} (for the bidisperse case).  

For the monodisperse case, only $\mathcal{D}_{ij}$ contributes 
to the relative velocity. At large Reynolds numbers, a $St^{1/2}$ 
scaling of the relative velocity in the inertial range is found for 
both the ballistic separation or the Richardson separation. 
Therefore, for the two-phase separation that well fits the 
simulation results, we have the same inertial-range 
scaling. This scaling is consistent with that from the differenical 
model by Zaichik and collaborators \cite[][]{zai03a, zai03, zai06}. 
Our model provides a clear physical picture for this scaling. 

      
Our calculations for the bidisperse case show that $\mathcal{A}_{ij}$ 
dominates the contribution to the relative velocity between particles 
of very different sizes, while for similar particles the primary 
contribution is from $\mathcal{D}_{ij}$. In the relative velocity vs. 
$St_2$ curves with fixed $St_1$, dips are found around $St_2 \sim St_1$,  
indicating stronger velocity correlation for similar size particles 
than for different size ones. 
Away from the dips, the relative velocity 
is essentially given by the contribution from $\mathcal{A}_{ij}$. 

The main assumptions in our model are those for the 
trajectory correlation and trajectory structure tensors. 
The approximations for these tensors can be tested 
and improved by numerical simulations, and our work thus 
provides a motivation for direct studies of these correlations 
along the particle trajectories. A direct numerical  
study of the separation behavior of particles backward in time
would also be of interest, because we have shown that it plays 
an important role  in modeling the relative velocity 
between particles of similar sizes. With the help of future simulations, 
the assumptions in our model could be considerably refined. The model may also 
be extended to include gravity and other effects.  
The refined and extended model would provide a 
reliable prediction of the relative velocity between inertial particles 
and can be applied to many practical studies, such as raindrop formation 
in atmospheric clouds and collisions of 
dust grains in astrophysical environments.

\begin{acknowledgments}  
LB acknowledges support from NASA grants 08-NAI5-0018  and NNX09AD10G.
\end{acknowledgments}

\bibliographystyle{jfm}
\bibliography{ms}

\end{document}